\documentclass[article,aps,prb,twocolumn,floatfix,showkeys,superscriptaddress,amssymb,preprintnumbers]{revtex4-1}
\usepackage[utf8]{inputenc}
\usepackage{amsmath}
\usepackage{amsfonts}
\usepackage{amssymb}
\usepackage{amsmath}
\usepackage{amsthm}
\usepackage{amstext} 
\usepackage{amsbsy}
\usepackage{graphicx}
\usepackage{textcomp}
\usepackage{color}
\definecolor{rot}{rgb}{1,0,0}
\definecolor{gruen}{rgb}{0,0.3,0}
\definecolor{normal}{rgb}{0,0,0}
\definecolor{blau}{rgb}{0,0,1}
\usepackage[T1]{fontenc}
\usepackage{graphicx}
\usepackage{mathrsfs}
\usepackage{paralist}
\usepackage{mathtools}
\usepackage{leftidx,tensor}
\usepackage{multirow}
\usepackage{rotating}
\usepackage{url}
\usepackage{hyperref}

\newcommand{\gb}{{\rm gb}}
\newcommand{\vv}{{\rm v}}

\begin{document}

\title{Grain boundary diffusion in CoCrFeMnNi high entropy alloy: \\kinetic hints towards a phase decomposition}

\author{Marcel Glienke}
\affiliation{Institute of Materials Physics, University of M\"unster, 48149 M\"unster, Germany}
\author{Mayur Vaidya}
\affiliation{Institute of Materials Physics, University of M\"unster, 48149 M\"unster, Germany}
\author{K. Gururaj}
\affiliation{Department of Metallurgical $\&$ Materials Engineering, Indian Institute of Technology; Madras, Chennai, 600036, India}
\author{Lydia Daum} 
\author{Bengü Tas}
\affiliation{Institute of Materials Physics, University of M\"unster, 48149 M\"unster, Germany}\author{Lukas Rogal}
\affiliation{Institute of Metallurgy $\&$ Materials Science, Polish Academy of Sciences, Krakow, Poland}
\author{K.G. Pradeep}
\affiliation{Department of Metallurgical $\&$ Materials Engineering, Indian Institute of Technology; Madras, Chennai, 600036, India}
\author{Sergiy V. Divinski}
\affiliation{Institute of Materials Physics, University of M\"unster, 48149 M\"unster, Germany}
\author{Gerhard Wilde}
\affiliation{Institute of Materials Physics, University of M\"unster, 48149 M\"unster, Germany}


\date{\today}

\begin{abstract}
Grain boundary self-diffusion of $^{57}$Co, $^{51}$Cr, $^{59}$Fe and $^{54}$Mn in a coarse-grained, single-phase fcc CoCrFeMnNi high entropy alloy is measured in a wide temperature range of 643 to 1273~K in both C- and B-type kinetic regimes after Harrison's classification. The results suggest that the product of the pertinent segregation factors, $s$, and the grain boundary width, $\delta$, is about 0.5~nm for all elements at temperatures $T>800$~K. Whereas one short-circuit contribution is observed at higher temperatures above 800~K, the penetration profiles in the C-type kinetic regime (643 -- 703~K) reveal two distinct contributions that hint towards a phase decomposition at a fraction of high-angle grain boundaries at these temperatures. A correlative microscopy combining transmission Kikuchi diffraction and atom probe tomography manifests formation of neighboring Ni-Mn-rich and Cr-rich precipitates at a segment of high angle grain boundaries. Transmission electron microscopy revealed an increased dislocation density in the vicinity of such interfaces which is suggested to be a reason of the enhanced diffusion rates at low temperatures for such short circuits. 

\end{abstract}

\maketitle

\section{Introduction}

High entropy alloys (HEAs), based on a novel design concept of multi-principal element combinations, contain equiatomic or near-equiatomic proportions of the constituent elements \cite{Yeh04}. These multicomponent alloys often show simple solid solution structures presumably stabilised by their high configurational entropy of mixing \cite{M14}. HEAs have shown potential both as structural and functional materials \cite{Chen19}. For example, Bhattacharjee et al. \cite{B18} have achieved an excellent combination of tensile strength (1562~MPa) and ductility (14\%), by producing a dual phase microstructure (L1$_2$ + B2) in AlCoCrFeNi$_{2.1}$ HEA. The equiatomic CoCrFeMnNi and CoCrFeNi alloys have shown increased strength and ductility at low temperatures due to twin formation \cite{Sun}. Al$_{0.6}$CoCrFeNi HEA exhibits three times the wear resistance than the conventional CGr15 steel at 873 K \cite{Chen19}, while exciting thermoelectric properties are exhibited by the Ti$_2$NiCoSnSb half-Heusler HEA \cite{K19}. 

'Sluggish' diffusion, in fact postulated originally as a core effect of HEAs \cite{Yeh04}, appears to be an ambiguous concept as it was shown in recent investigations utilizing the radiotracer methods \cite{Ni, all, Josua}, interdiffusion \cite{Mayur-ther} and tracer-interdiffusion measurements \cite{Daniel}. The experimental observations of diffusion-controlled processes in HEAs have also hinted towards a non-sluggish atomic transport in these multicomponent alloys \cite{Chokshi, Mayur-grow}. 

However, nearly all the diffusion investigations in HEAs so far (except the very first report on Ni grain boundary diffusion in CoCrFeMnNi by Vaidya et al. \cite{GB}), have been limited to bulk diffusion. It is well established that grain boundaries (GBs) offer high-diffusivity paths in metals and alloys \cite{Aloke-book}. GB diffusion contributes to or even dominates a number of important phenomena such as Coble creep, recrystallization, grain growth, sintering, diffusion-induced GB migration and various discontinuous reactions \cite{KMG}. GB diffusion coefficients have also served as key inputs in computational techniques to develop models for hot forming processes \cite{N14} and to investigate creep behaviour in nanocrystalline materials at the atomic scale \cite{B15}. GB diffusion measurements are a sensitive tool, too, for the determination of the segregation behaviour of solutes in various alloy systems from a dilute solid solution  to concentrated alloys \cite{Henning, nl-seg, Esin}. With HEAs being continuously explored for various applications utilizing their unique feature of no-solute and no-solvent matrix, investigation of the corresponding GB diffusion rates is of high fundamental as well as technological relevance. Furthermore, prolonged annealing treatments \cite{Otto} or plastic deformation \cite{creep} were reported to induce a phase decomposition in CoCrFeMnNi HEA, especially at grain boundaries. 

This motivated us to perform a detailed investigation of GB diffusion in a CoCrFeMnNi HEA, which in fact may be considered as a model five-component high-entropy alloy, with the following objectives:

\begin{itemize}
\item Determination of GB self-diffusion coefficients of Co, Cr, Fe and Mn in CoCrFeMnNi that in addition to the available data on Ni \cite{GB} will provide a complete data set for this classical HEA;
\item Evaluation of the impact of potential phase decomposition (nano-scale precipitation) at moderate and low temperatures on GB diffusion.
\end{itemize}

In our recent study \cite{GB}, grain boundary self-diffusion coefficients of Ni in CoCrFeNi and CoCrFeMnNi HEAs were determined using the radiotracer analysis. A non-sluggish diffusion behaviour was observed when Ni GB diffusivities in CoCrFeMnNi HEA were compared with other FCC alloys. A cross-over absolute (at about 800~K) or homologous ($0.46T_m$) temperature was found to exist above which Ni GB diffusion in 5-component CoCrFeMnNi HEA became enhanced with respect to that in 4-component CoCrFeNi. ($T_m$ is the melting point of the corresponding compound.) This observation highlighted the need for determining the grain boundary self-diffusion behaviour for all the other constituents. 

The CoCrFeMnNi HEA initially was believed to be one of the most stable HEAs. However, as it was already mentioned, recent findings have indicated precipitate formation in this alloy either after severe plastic deformation followed by an annealing treatment or after very long annealing times at a low temperature of 773~K \cite{nc-tol, Lapl, Otto}. In the latter case, Otto et al. \cite{Otto} documented that micron-sized precipitates, identified as $L1_0$ Ni-Mn, B2 Co-Fe and BCC Cr-rich particles, are formed at grain boundaries. A nano-scale precipitation was observed in CoCrFeMnNi during creep at intermediate temperatures of 783--856~K \cite{creep}. Such nano-scale precipitation at GBs can change the local defect structures near the boundaries, without significantly affecting the GB crystallography. One may expect precipitation in multi-component alloys even at low-angle GBs, too, as it was observed for dislocation pipes \cite{Turlo-Rup}.

Therefore, the measurements of GB self-diffusion of constituent elements can be applied as a sensitive probe of nano-precipitation at the interfaces providing an integrated information over the whole sample volume. In the present study, we combined these experiments with a (local) correlative microscopic analysis of the CoCrFeMnNi HEA annealed at 673~K (the same conditions as for the diffusion measurements were used) in order to characterize general high-angle grain boundaries in detail.

\section{Experimental details}

\subsection{ \emph{Alloy preparation and characterization}}
The equiatomic CoCrFeMnNi alloy was produced by high-frequency induction melting of metal pieces of Co, Cr, Fe, Mn and Ni of $99.9 \%$ purity. The samples were homogenized at $1373$~K for 50~h, polished and pre-annealed at the diffusion measurement temperatures for the intended diffusion annealing times (details given later). This treatment ensured equilibrium segregation at grain boundaries and reduces mechanical stresses induced into the sample by polishing. 

The microstructure of the pre-annealed alloys was investigated using X-ray diffraction (XRD) with Cu K$_{\alpha}$ radiation (measured with a step size of $0.02 ^\circ$ for 2 s) and scanning electron microcopy (SEM, FEI Nova NanoSEM 230) equipped with electron back-scatter diffraction (EBSD, TSL-OIM software) and energy dispersive X-ray spectroscopy (EDS). 

The local chemical homogeneity of the alloys was further characterized using a local electrode atom probe tomography (LEAP 5000X HR) provided by CAMECA instruments. Site specific tips containing a chosen high-angle GB for atom probe tomography (APT) was prepared using a FEI Helios Nanolab G4 UX dual beam focussed ion beam following the procedures described in Refs. \cite{Pradeep1, Pradeep2}. APT measurements were performed with the tips maintained at $60$ K applying laser pulses at $250$ kHz frequency and $30$ pJ laser energy. Data reconstruction and analysis was performed with an IVAS 3.8.4 software. 

\subsection{\emph{Radiotracer measurements}}

The energies of $\gamma$-decays and half-time of the radioactive isotopes used are given in Table~\ref{tab:gamma}.

\begin{table}[ht]
\caption{Half-time and energies of $\gamma$-decays of the radioactive isotopes used in the present investigation.}\label{tab:gamma}
\begin{center}
\begin{tabular}{c|cc}
	\hline
	Isotope & Half-time (days) & Energy (keV) \\
	\hline
	$^{57}$Co & 271.7 & 122 \\
	$^{51}$Cr & 27.7  & 320 \\
	$^{54}$Mn & 312   & 834 \\
	$^{59}$Fe & 44.6  & 1095 \\
	\hline
\end{tabular}
\end{center}
\end{table}

The radioactive tracers, available as highly diluted acidic solutions, were mixed and applied to the surface of the disc-shaped, pre-annealed samples of $7.5$~mm diameter and $1$~mm thickness, which were mechanically polished to a mirror finish. Since we applied tiny amounts of the isotopes already available as chemical elements in the alloy, this procedure does not change chemical equilibrium in the sample. Thus, true self-diffusion behavior was measured without inducing undesirable co-segregation effects. 

After tracer application, the samples were sealed in quartz tubes under purified (5N) Ar atmosphere and subjected to annealing treatments at the temperatures of 643, 673, 703~K (i.e. the C-type kinetic regime measurements) and 973, 1073, 1173, 1273~K (experiments in the B-type kinetic regime) for appropriate times. For both pre-annealing and diffusion annealing, the furnace temperatures were controlled within $\pm$1K using Ni/NiCr thermocouple (type K). 

After annealing the samples were quenched in water, followed by reducing the diameter of the samples by roughly 1 mm. This was done to avoid any effect due to lateral diffusion. 

A high precision grinding machine was used to measure the penetration profiles by sectioning thin slices with an accuracy of $\pm0.05$~$\mu$m. The relative activity of each section was measured using a solid Ge-detector equipped with a 16K multi-channel energy discriminator (energy resolution of about 0.7~keV), which enabled a differentiation of the activities of all the four tracers, see Table~\ref{tab:gamma} for the corresponding energies. Thus, the penetration profiles for all the tracers could be determined in a single experiment at a given temperature. The counting times of the detector were chosen to keep the statistical uncertainties below 2$\%$.

\section{Results}

\subsection{Microstructure}

\begin{figure}[ht]
\centering 	\includegraphics[width=0.9\linewidth]{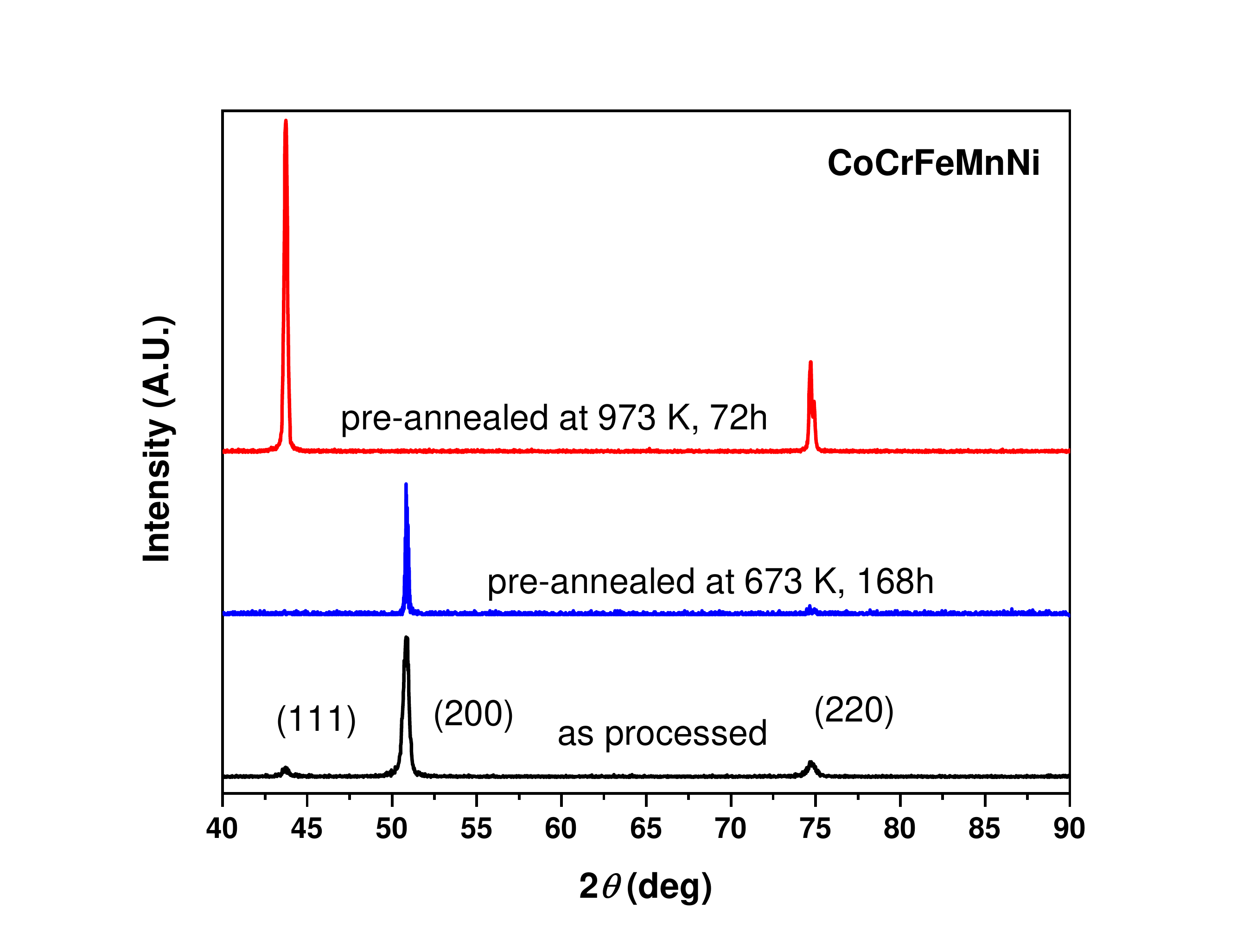}
		\caption{XRD-pattern of the CoCrFeMnNi, pre-annealed at the conditions relevant for diffusion measurements. All peaks occurring in the spectrum refer to a single-phase FCC structure.}
\label{fig:XRD}
\end{figure}

XRD analysis was used to confirm a single-phase structure for the CoCrFeMnNi HEA in the as-processed state and its stability during diffusion annealing. Figure~\ref{fig:XRD} shows the X-ray spectra for the as-cast state and two further samples annealed at 673~K for 168~h and at 973~K for 72~h. The results confirm that a single-phase FCC structure is macroscopically retained during diffusion annealing treatments for all the samples. Note that not all Bragg reflections could simultaneously be recorded due to the large grain size that prohibits reliable estimations of the texture and its evolution with the annealing treatment. However, this information is not relevant for the intended diffusion studies (diffusion in crystals with cubic symmetry is isotropic).

Orientation imaging microscopy reveals a coarse-grained microstructure with an average grain size in excess of 200~$\mu$m, see the EBSD maps presented in Fig.~\ref{fig:EBSD}a as an example. This ensures the correct kinetic conditions for the diffusion measurements, see below. Although the performed EBSD measurements are statistically not reliable for accurate grain size determination, the actual value of the grain size does not affect the final results. Figure~\ref{fig:EBSD}b presents strain maps obtained using the kernel average misorientation substantiating the absence of significant stresses in the bulk of the sample. The corresponding elemental distribution shown in Fig.~\ref{fig:EBSD}c confirms a homogeneous distribution of the constituting elements and an absence of any micro-scale phase separation.

\begin{figure*}[t]
\centering \includegraphics[width=0.99\textwidth]{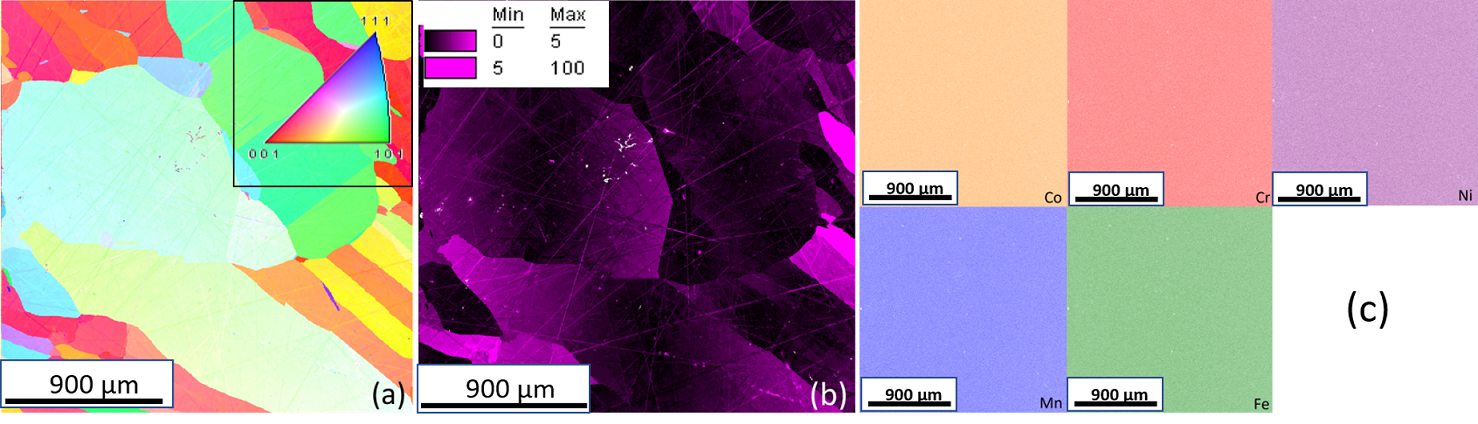}
\caption{An example of EBSD pattern (a),  kernel average misorientation (b) and chemical element distribution (c) in the CoCrFeMnNi alloy annealed at 673~K for 168~h. Inverse pole figure color mapping, see insert, is used for the EBSD pattern in (a).}
\label{fig:EBSD}
\end{figure*}

\subsection{Radiotracer diffusion measurements}

Analysis of the GB diffusion profiles depends strongly on the pertinent kinetic conditions \cite{KMG}. Harrison \cite{Har} classified three different kinetic regimes (types A, B and C) from which the B- and C-type kinetics are relevant for the present study. In these cases, different approximations of the exact solution of the Fisher problem \cite{Fisher} given by Suzuoka \cite{Suz} and Whipple \cite{Wip} should be used, see the extended discussion in Ref. \cite{KMG}. 

\subsubsection{B-type kinetic regime}

The B-type regime holds if the value of the Le Claire parameter \cite{LeClair} $\alpha$,
\begin{equation}\label{eq:alpha}
\alpha = \frac{s \delta }{2 \sqrt{D_\vv t}},
\end{equation}
is smaller than $0.1$ and $\Lambda > 3$ \cite{Aloke-book}. The parameter $\Lambda$ is determined as \cite{Aloke-book}
\begin{equation}\label{eq:Lambda}
\Lambda = \frac{d}{\sqrt{D_\vv t}} .
\end{equation}
In these expressions $d$ is the grain size, $t$ the diffusion time, $D_v$ the volume diffusion coefficient, $s$ the segregation factor, and $\delta$ the GB width. These two conditions guarantee that the volume diffusion length, $\sqrt{D_\vv t}$, is larger than the effective GB width, $s\times\delta$, but smaller than the grain size, $d$. Thus, the diffusion fluxes originating from different GBs do not overlap in the crystalline bulk. 

The solute segregation factor, $s$, is generally defined as the ratio of the solute concentration at a GB, $c_\gb$, to that in the adjacent crystalline bulk, $c_\vv^0$, $s=c_\gb/c_\vv^0$ \cite{Herzig}. In the present case of a multi-principal element alloy, both tracer and constituting atoms of the same chemical element contribute to the thermodynamic equilibrium at the interface \cite{Esin}. Bernardini with co-workers \cite{Bern1, Bern2} have proven that an equilibrium segregation is established shortly after commencement of a B-type diffusion measurement and the total concentration of the given chemical element has to be taken into account to determine the value of the corresponding segregation factor $s$. Since the tracer concentration is typically very small in a radiotracer experiment, it can safely be neglected and the chemical element concentrations at a GB and in the bulk of the alloy determine the equilibrium segregation factor \cite{nl-seg}. Thus, one should expect small values of the segregation factors, about unity in the present case. For example, $s$ would be about 5 for Ni in a hypothetical case when a monolayer of pure Ni would cover the GBs in the HEA. Since no enrichment of any element was seen in the CoCrFeMnNi alloy at elevated temperatures of about 1000~K \cite{GB}, the value of $s$ of about unity can safely be used for initial estimates.

The diffusional GB width $\delta$ was \emph{measured} in dedicated GB diffusion experiments in a number of fcc metals and alloys and the value of $\delta=0.5$~nm was found to be a reliable estimate \cite{Dasha, Ger}. In this paper we will show that this value holds for the CoCrFeMnNi HEA, too, see below.

The B-type regime corresponds to moderate annealing times and temperatures. In this regime both bulk and GB diffusion occur simultaneously, but the GB diffusion coefficient is significantly larger. As a result, two distinct diffusion branches appears and bulk and GB contributions could reliably be separated. 

\begin{figure*}[t]
\centering \includegraphics[width=0.9\textwidth]{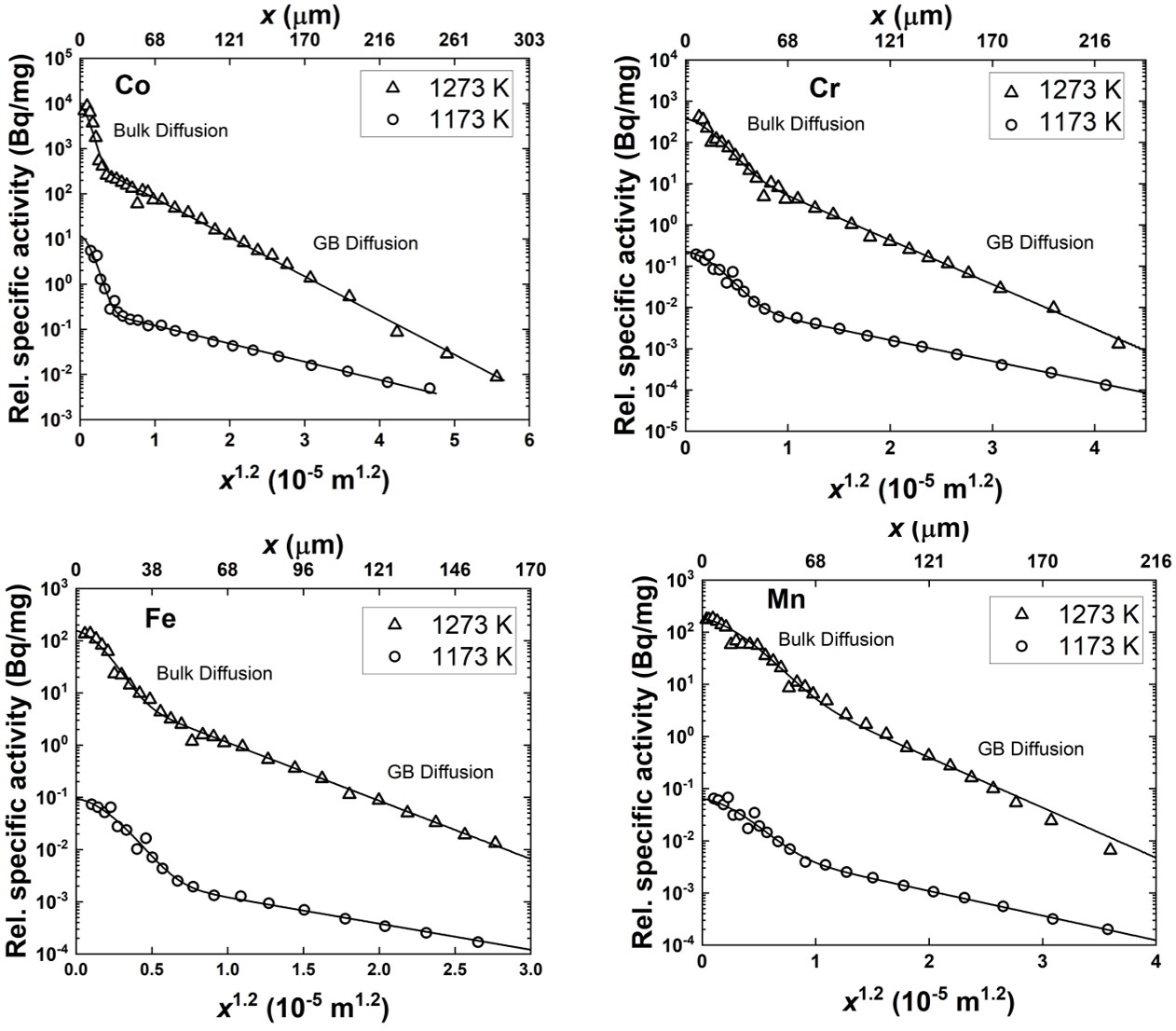}
\caption{Penetration profiles determined in the B-type kinetic regime at 1173 K and 1273 K for Co (a), Cr (b), Mn (c) and Fe (d). The  profiles measured at 1173~K have been re-scaled for all tracers multiplying the depths to the 6/5$^{\rm th}$ power by a factor of $4$ and the relative specific activities by a factor of $0.001$ for a better visualization.}
\label{fig:B-kinetic_profiles}
\end{figure*}

Typical penetration profiles measured at 1273~K and 973~K in the B-type regime are shown in Fig.~\ref{fig:B-kinetic_profiles}. The near-surface branch of the profiles corresponds to bulk diffusion and follows Gaussian behavior for the instantaneous source initial conditions \cite{Aloke-book}. The deeper branch of the diffusion profile corresponds to grain boundary diffusion and follows approximately the $\ln C \sim x^{1.2}$ dependency \cite{LeClair}. Thus, the penetration profiles were fitted as a sum of two contributions,
\begin{equation} \label{eq:Fitting_function}
C = A_\vv \exp \big(- B_\vv x^{2} \big)+ A_\gb \exp \big(- B_\gb x^{1.2} \big),
\end{equation}
with $A_\vv$, $B_\vv$, $A_\gb$, and $B_\gb$ being the corresponding fit parameters. According to Le Clair’s analysis of Whipples and Suzuoka exact solutions \cite{LeClair}, it is the  triple product $P$,
\begin{equation}\label{eq:P_value}
P = s \delta D_{gb} = 1.322 \sqrt{\frac{D_v}{t}} \bigg(- \frac{\partial \text{ln} C}{\partial x^{6/5}} \bigg)^{-5/3},
\end{equation}
which can be determined from the corresponding slope of the GB diffusion-related branch of the penetration profile plotted in the coordinates of $\ln C$ vs. $x^{1.2}$. The approximation (\ref{eq:P_value}) holds if the second Le~Clair parameter $\beta$ \cite{LeClair}, 
\begin{equation}\label{eq:beta}
\beta = \alpha \left( \frac{D_\gb}{D_\vv} -1 \right) \approx \frac{P}{2D_\vv \sqrt{D_\vv t}} ,
\end{equation}
is large enough, $\beta>10^4$ \cite{KMG}. Slightly changed expressions have to be used instead of Eq.~(\ref{eq:P_value}) at smaller values of $\beta$.

Bulk diffusion coefficients, $D_\vv$, are required to determine the triple product $P$ according to Eq.~(\ref{eq:P_value}). In the present work, the bulk diffusion coefficients measured for CoCrFeMnNi single crystals \cite{Daniel-sc} are used. The experimental parameters and the determined triple products are listed in Table~\ref{tab:B-Param}. 

Some profiles were measured under conditions for which $\beta<10$. In such cases we applied the exact Suzuoka solution \cite{Suz} and determined the triple products. The corrections were typically not large and they do not exceed 15\% in the present case. If $\beta>10$, the approximate Le Clair solutions provide results of sufficient accuracy with respect to the experimental uncertainties. The latter do not exceed 10\% in the present work.

\begin{table*}[t] 
\caption{Experimental parameters (temperature $T$ and diffusion time $t$) and the determined triple products $P$ for the B-type kinetic measurements. The parameters $\alpha$, $\beta$ and $\Lambda$ were determined by Eqs.~(\ref{eq:alpha}), (\ref{eq:beta}), and (\ref{eq:Lambda}), respectively. For the small values of $\beta$, $\beta < 10$ (marked with asterisks) the exact Suzuoka solution \cite{Suz} was used instead of Eq.~(\ref{eq:P_value}).}
\label{tab:B-Param}
\begin{center}
\begin{tabular}{c | c | c | c | c | c | c | c }
	
	$T$  & $t$ & \multirow{2}{1.1cm}{Tracer} & $\sqrt{D_v t}$  & $P$   & $\alpha$ & \multirow{2}{1.1cm}{~~~$\beta$} & \multirow{2}{1.1cm}{~~~$\Lambda$} \\
	(K)& ($10^3$ s) & & ($\mu$m)& (m$^3$/s) & $10^{-4}$ & & \\   
	\hline 
	& & & & & & & \\
	$973$ & $72$ & Co & $0.25$ & $(6.13 ^{+3.5}_{-1.7}) \times 10^{-22}$ & $10$ & $5340$ & $407$ \\[4pt]
	&  & Cr & $0.22$ & $(3.35^{+1.8}_{-0.99}) \times 10^{-22}$ &$12$ & $4233$ & $460$ \\[4pt]
	&  & Mn & $0.45$ & $(2.99 ^{+2.1}_{-1.3} ) \times 10^{-21}$ & $5.5$ & $4166$ & $220$ \\[4pt]
	&  & Fe & $0.31$ & $(2.21^{+1.4}_{-1.1}) \times  10^{-22}$ & $8.2$ & $994$ & $326$\\[4pt] \hline
	& & & & & & & \\[-4pt]
	$1073$ & $72$ & Co & $0.96$ & $(1.61^{+0.4}_{-0.2} ) \times  10^{-21}$ & $2.6$ & $237$ & $104$ \\[4pt]
	&  & Cr & $1.08$ & $(1.25^{+0.48}_{-0.21} ) \times  10^{-21}$ & $2.3$ & $129$ & $93$ \\[4pt]
	&  & Mn & $1.99$ & $(5.07^{+2.1}_{-1.07} ) \times  10^{-21}$ & $1.3$ & $84$ & $50$ \\[4pt]
	&  & Fe & $1.27$ & $ (1.52^{+1.3}_{-1.1}  ) \times  10^{-21}$ & $1.9$ & $95$ & $79$\\[4pt]  \hline
	& & & & & & & \\[-4pt]
	$1173$ & $24$ & Co & $1.71$ & $(1.361^{+0.48}_{-0.23} ) \times 10^{-20}$ & $1.5$ & $118$ & $104$ \\[4pt]
	&  & Cr & $2.36$ & $(1.18^{+0.11}_{-0.08} ) \times 10^{-20}$ & $1.1$ & $129$ & $39$ \\[4pt]
	&  & Mn & $3.90$ & $(2.12 ^{+0.98}_{-0.45} ) \times 10^{-20}$ & $0.64$ & $84$ & $15$ \\[4pt]
	&  & Fe & $2.40$ & $(1.24^{+0.56}_{-0.31} ) \times 10^{-20}$ & $1.0$ & $95$ & $39$\\[4pt] \hline
	& & & & & & & \\[-4pt]
	$1273$ & $24$ & Co & $4.42$ & $(5.21^{+1.3}_{-0.8} ) \times 10^{-20}$  & $0.33$ & $15$ & $13$ \\[4pt]
	&  & Cr & $7.23$ &$(8.85^{+2.4}_{-1.13} ) \times 10^{-20} $ & $1.9$  & $5^*$ & $8$\\[4pt]
	&  & Mn & $10.97$ & $(9.34 ^{+0.6}_{-0.5} ) \times  10^{-20}$ & $0.15$  & $2.6^*$ & $5$ \\[4pt]
	&  & Fe & $6.49$ &  $(4.06^{+1.2}_{-0.98}  ) \times 10^{-20}$ & $0.28$ & $7.6^*$ & $9$\\[4pt]  
\end{tabular}
\end{center}
\end{table*}

\subsubsection{C-type kinetic regime}\label{sec:C}

\begin{figure*}[t]
\begin{center}
\includegraphics[width=0.9\textwidth]{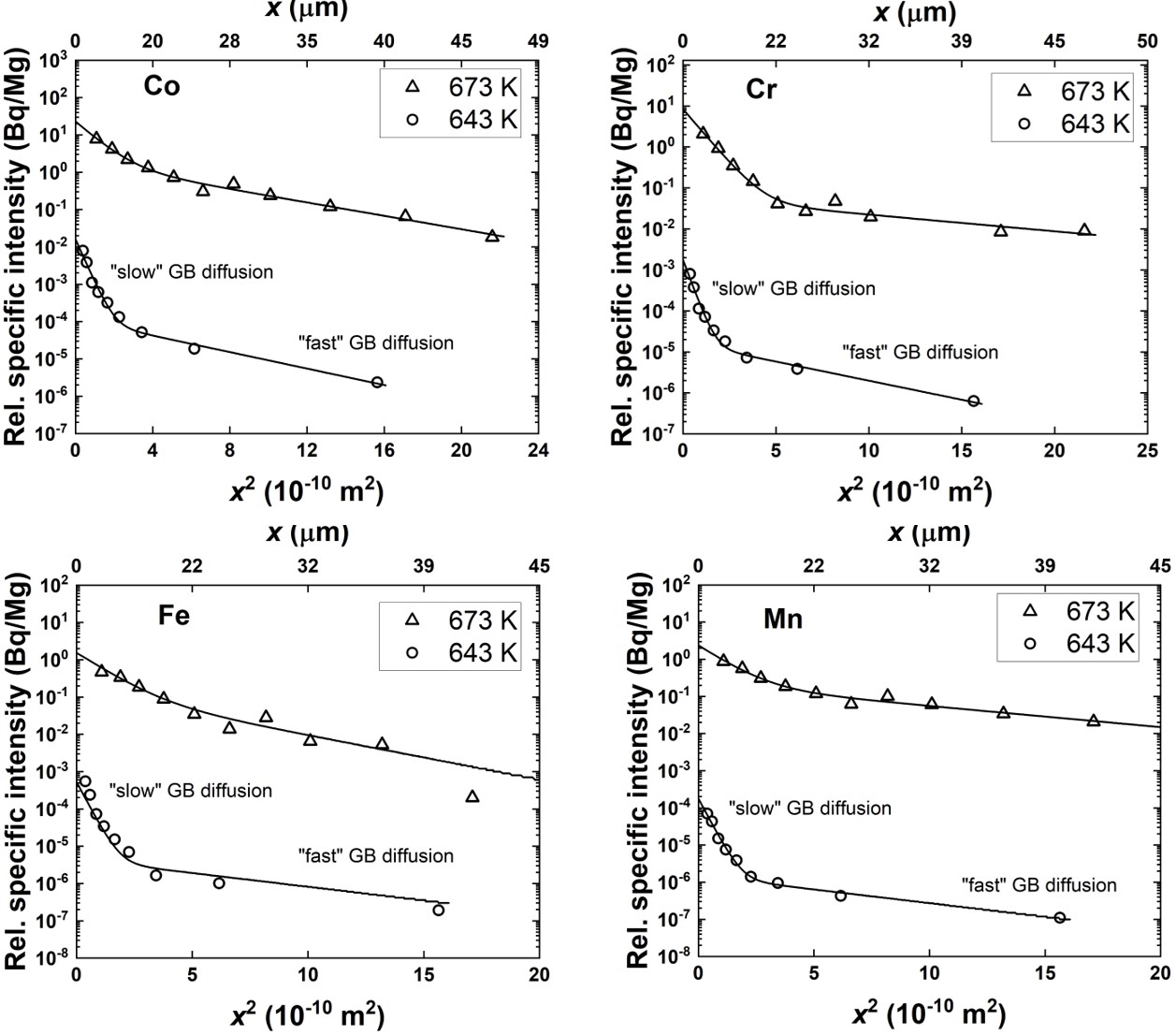}
\end{center}
\caption{Penetration profiles measured in the C-type kinetic regime at 643 and 673~K for Co (a), Cr (b), Mn (c) and Fe (d). For a better visibility, the 643~K profiles have been scaled for all tracers multiplying the $x^2$ values by a factor of $4$ and the relative specific activities by $0.001$ for a better visualization.}
\label{fig:C-kinetic_profiles}
\end{figure*}

The C-type kinetic regime corresponds to low temperatures and shorter annealing times, $\alpha > 1$ \cite{Aloke-book}. The value of the $\alpha$ parameter in this regime indicates that the bulk diffusion depth is much smaller than the effective grain boundary width $s\times \delta$. Typically this means that bulk diffusion is effectively frozen in this regime. Thus, the GB diffusion coefficient, $D_{\gb}$, can directly be determined using the Gaussian solution function for the instantaneous diffusion source problem \cite{Aloke-book},
\begin{equation}\label{eq:Gaus}
C = \frac{A}{\sqrt{D_\gb t}} \exp \left(- \frac{x^2}{4D_{\gb}t} \right).
\end{equation} 
Here $A$ is a fitting constant. The penetration profile for Co, Cr, Fe and Mn measured at 643 and 673~K are shown in Figs.~\ref{fig:C-kinetic_profiles}a, b, c, and d, respectively. Unexpectedly, all the profiles reveal two distinct short-circuit contributions for depths larger than 5~$\mu$m. Some artifacts for very near-surface sections at depths below few micrometers may be expected for the applied mechanical sectioning procedure and these points are not included in the analysis. Thus, with the expected experimental limitations, it can be  concluded that at least several distinct short-circuit fluxes with significantly different diffusion coefficients appear in the material during annealing at lower temperatures of about 650--700K. This fact suggests that distinct types of high-angle GBs appear to be in the alloy. Since the measurements were definitely performed in the C-type kinetic regime, see Table~\ref{tab:C-param}, we used a sum of two Gaussian-type solutions to fit the profiles with corresponding pre-factors,
\begin{equation}\label{eq:fitC2}
C = A_1 \exp\left(- \frac{x^2}{4D_{\gb,1}t} \right) + A_2 \left(- \frac{x^2}{4D_{\gb,2}t} \right),
\end{equation}
and thus yielding two diffusion coefficients (tentatively we assumed here that these are two GB diffusion coefficients, as it will be verified below) for each profile and tracer. 

If $0.1 <\alpha < 1$, the measurements fall into a so-called B-C-type transition regime \cite{Aloke-book}. This occurred to be the case for the profiles measured at 703~K. A correction procedure described in Ref.~\cite{Aloke-book} was used and the results are given in Table~\ref{tab:C-param}.

\begin{table*}[h] 
\caption{Experimental parameters (temperature $T$ and time $t$) of the C-type kinetic measurements and the determined GB diffusion coefficients $D_{\gb,1}$ and $D_{\gb,2}$ (see Eq.~(\ref{eq:fitC2})). The cases with $\alpha < 1$ (marked by asterisks) correspond to a B-C-type transition regime and they were analysed accordingly.}
\label{tab:C-param}
\begin{center}
\begin{tabular}{c | c | c | c | c | c | c}
	$T$ & $t$ & \multirow{2}{1.1cm}{Tracer} & $\sqrt{D_\vv t}$ & $D_{\gb,1}$ ('slow') & $D_{\gb,2}$ ('fast') & \multirow{2}{1cm}{~~~$\alpha$} \\
	(K) & (s)&  & ($10^{-10}$~m) & (m$^2$/s) & (m$^2$/s) & \\
	\hline  & & & & & & \\[-4pt]
	$643$ & $864000$ & Co & $0.51$ & $(4.29 ^{+3.7}_{-1.5} )\times 10^{-18}$ & $(3.25 ^{+7.5}_{-0.38})\times 10^{-17}$ &$4.4 $ \\[4pt]
	&  & Cr & $0.54$ & $(4.21^{+ 1.1}_{-0.35})\times 10^{-18}$ & $ (3.85^{+2.4}_{-1.1})\times 10^{-17}$ &$4.2 $ \\[4pt]
	&  & Mn & $0.39$ & $ (2.69^{+1.5}_{-0.27})\times10^{-18}$ & $(4.26 ^{+2.1}_{-0.18})\times 10^{-17}$ &$6.4 $ \\[4pt]
	&  & Fe & $0.47$ & $(2.70^{+1.3}_{-1.0})\times 10^{-18}$ & $(3.87 ^{+2.4}_{-0.5} )\times 10^{-17}$ &$5.3 $  \\[4pt]
	\hline  & & & & & & \\[-4pt]
	$673$ & $604800$ & Co & $6.4$ & $(3.86^{+1.4}_{-1.1} )\times10^{-17}{}$ & $(2.00 ^{+2.3}_{-1.3})\times10^{-16}$ &$1.7 $ \\[4pt]
	&  & Cr & $1.26$ & $(3.38^{+1.2}_{-0.9} )\times10^{-17}$ & $(4.41^{+3.4}_{-1.2})\times 10^{-16}$ &$1.9 $ \\[4pt]
	&  & Mn & $1.67$ & $(4.32 ^{+2.3}_{-1.2} )\times10^{-17}$ & $(3.15 ^{+2.4}_{-1.0} )\times10^{-16}$ &$1.5$ \\[4pt]
	&  & Fe & $1.50$ & $(4.16^{+4.1}_{-1.5} )\times10^{-17}$ & $(1.49^{+2.5}_{-0.9} )\times 10^{-16}$ &$1.7 $  \\[4pt]
	\hline  & & & & & & \\[-4pt]
	$703$ & $86400$ & Co & $5.24$ & $(9.80^{+4.5}_{-2.2} )\times10^{-17}$ & $(2.66 ^{+2.2}_{-1.0} )\times10^{-15}$ & $0.5^* $ \\[4pt]
	&  & Cr & $1.70$ & $(1.50^{+0.7}_{-0.2} )\times10^{-16}$ & $(1.2^{+0.6}_{-0.2} )\times10^{-15}$ &$1.4 $ \\[4pt]
	&  & Mn & $5.93$ & $(1.85 ^{+0.9}_{-0.3} )\times10^{-16}$ & $(3.24^{+0.6}_{-0.15} )\times 10^{-15}$ &$0.4^* $ \\[4pt]
	&  & Fe & $4.99$ & $(7.50^{+1.15}_{-0.4} )\times10^{-17}$ & $(3.19 ^{+3.0}_{-1.6} )\times10^{-15}$ &$0.5^* $ \\[4pt]
\end{tabular} 
\end{center}
\end{table*}

\subsection{Temperature dependence of grain boundary diffusion}
\begin{figure*}[ht]
\begin{center}
\includegraphics[width=0.9\textwidth]{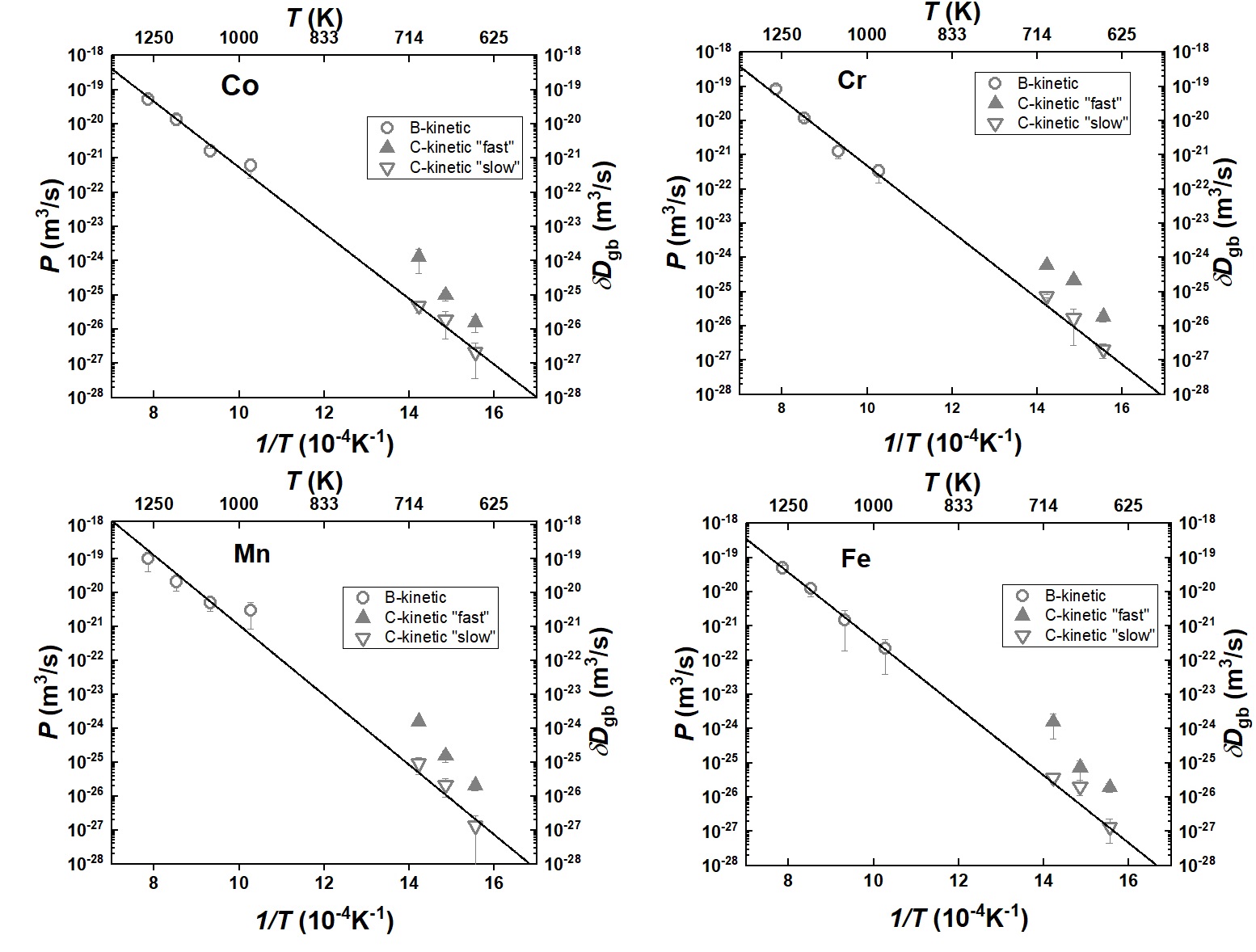}
\end{center}
\caption{Arrhenius plots for diffusion of Co (a), Cr (b), Mn (c) and Fe (d) determined in the B- (the triple products $P$, open circles, left ordinate) and C-type (the products of the measured diffusion coefficients $D_{\gb,1}$, $D_{\gb,2}$ and of the GB width $\delta=0.5$~nm, triangles, right ordinate) kinetic regimes.}
\label{fig:Arrprofiles}
\end{figure*}

Since the value of the GB width $\delta=0.5$~nm was found to be a good estimate \cite{Dasha}, we will use it in the analysis. In Fig.~\ref{fig:Arrprofiles}, the temperature dependencies of the triple products (measured in the B-type kinetic regime) and of the products of the GB diffusion coefficients (measured in the C-type kinetics) and the GB width $\delta$, i.e. $\delta \times D_{\gb,1}$ and $\delta \times D_{\gb,2}$, are plotted together. Common Arrhenius-type temperature dependencies, solid lines, are found for the $P$-values and the products $\delta \times D_{\gb,1}$, Fig.~\ref{fig:Arrprofiles}, while the values of $\delta \times D_{\gb,2}$ lie consistently higher.

\section{Discussion}
The XRD measurements and the EBSD patterns, as well as the EDX maps confirm a coarse-grained, single-phase fcc microstructure, at least down to a micrometer scale. The different intensities of the peaks visible in the XRD measurements (Fig.~\ref{fig:XRD}) could correspond to a texture evolution during the annealing treatment. However, the EBSD measurements performed on the as-cast and annealed samples show an average grain size in excess of 200~$\mu$m. It is thus more probable that the X-ray beam was focused on a few grains and the XRD intensities were affected by re-mounting the sample after annealing. Since diffusion in cubic materials is isotropic, a potential texture evolution cannot affect the measured diffusion rates and the conclusions of the present work.

\subsection{Temperature dependence of grain boundary diffusion}

In Figs.~\ref{fig:Arrprofiles}a--d, the GB triple products, $P$, obtained from the B-type kinetic measurements and the products $\delta \times D_{\gb,1}$ and $\delta \times D_{\gb,2}$ determined from the C-type kinetic data are plotted as a function of the inverse absolute temperature $T$. Here we are using the same value of the GB width, $\delta$, for the two types of interfaces, $\delta=0.5$~nm. In general, the diffusional GB widths should not be the same for the two types of boundaries, but we are not expecting a large difference and it is probably limited by a factor of two to three. Note that the structure width of 'deformation-modified' grain boundaries induced by severe plastic deformation was estimated at 1.5 to 2~nm \cite{Xav, neGBs} and this is probably an upper limit for the thickness of the interface region with significantly enhanced diffusion coefficients. The approximation $\delta=0.5$~nm is reasonable in view of the logarithmic scale used in Fig.~\ref{fig:Arrprofiles}.

The values of $P$ and $\delta \times D_{\gb,1}$ fall systematically on the same Arrhenius line for all isotopes and correspond to diffusion along random high-angle GBs in the alloy. Since these results imply that $s\times \delta=0.5$~nm, one can determine the corresponding Arrhenius parameters, namely the pre-exponential factor $P_0$ and the activation enthalpy $Q$,
\begin{equation}\label{eq:Arrh}
P(T) = P_0 \exp \left(-\frac{Q}{RT} \right) ,
\end{equation}
and the results are given in Table~\ref{tab:Arrhparam}.

\begin{figure}[ht]
	\begin{center}
		\includegraphics[width=0.9\linewidth]{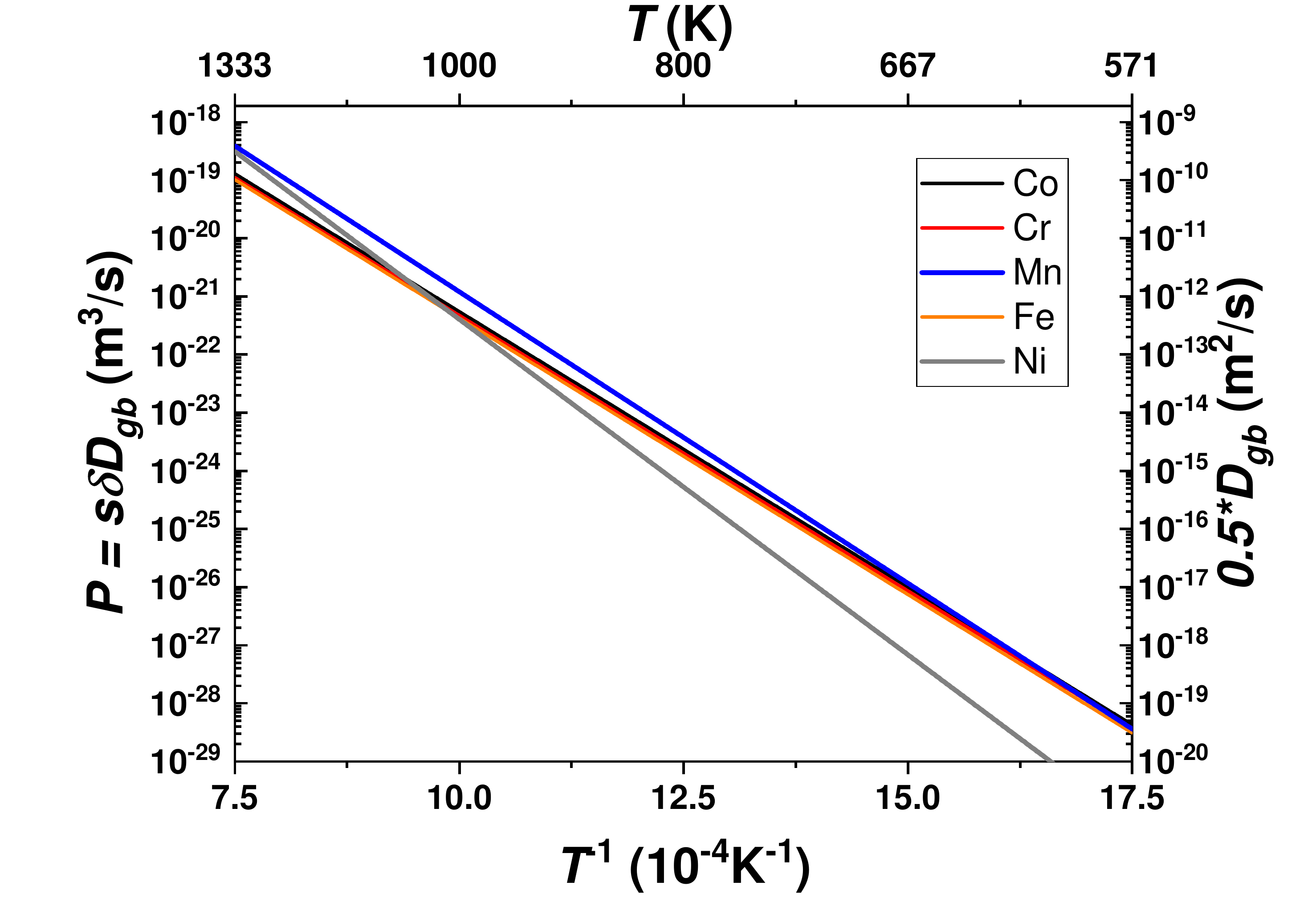}
	\end{center}
	
	\caption{GB diffusion rates of constituent elements in CoCrFeMnNi (the Ni data are taken from Ref.~\cite{GB}).}
	\label{fig:Comp_other_fcc1}
\end{figure}

Figure~\ref{fig:Comp_other_fcc1} compares the Arrhenius plots of Co, Cr, Fe, Mn and Ni tracers on the inverse absolute temperature scale. The individual elements seem to have roughly the same diffusivities and the Arrhenius parameters, see Table~\ref{tab:Arrhparam}, with only Mn and Ni showing minor deviations. The proximity of GB self-diffusivities of all the constituents in the CoCrFeMnNi HEA resembles that for their bulk diffusion coefficients \cite{all}.

\begin{table}[ht]
	\caption{Arrhenius parameters for GB diffusion of the constituting elements in CoCrFeMnNi HEA. The data for Ni, marked by asterisk, are taken from Ref.~\cite{GB}.}
	\label{tab:Arrhparam}
	\begin{center}
		\begin{tabular}{c | c | c}
			\hline
			\multirow{2}{1.1cm}{Tracer} & $P_0$ & $Q$ \\ 
			& ($10^{-12}$ m$^3$/s) & (kJ/mol) \\
			\hline
			& & \\[-4pt]
			Co & $1.66_{-0.79}^{+1.5}$ & $181.5 \pm 4.5$  \\[4pt]
			Cr & $1.43_{-0.66}^{+1.2}$ & $180.6 \pm 4.3$    \\ [4pt]
			Mn & $13.0_{-9.3}^{+33}$ &  $192.1 \pm 8.9$ \\[4pt]
			Fe & $1.40_{-0.64}^{+1.40}$ &  $182.2 \pm 4.3$  \\[4pt]
			Ni$^*$ & $ 142_{-123}^{+915}$ & $221 \pm 14$  \\[4pt]
			\hline
		\end{tabular}
	\end{center}
\end{table}

\subsection{Comparison of the diffusivities with other FCC matrices}

\begin{figure}[ht]
	\begin{center}
		\includegraphics[width=0.8\linewidth]{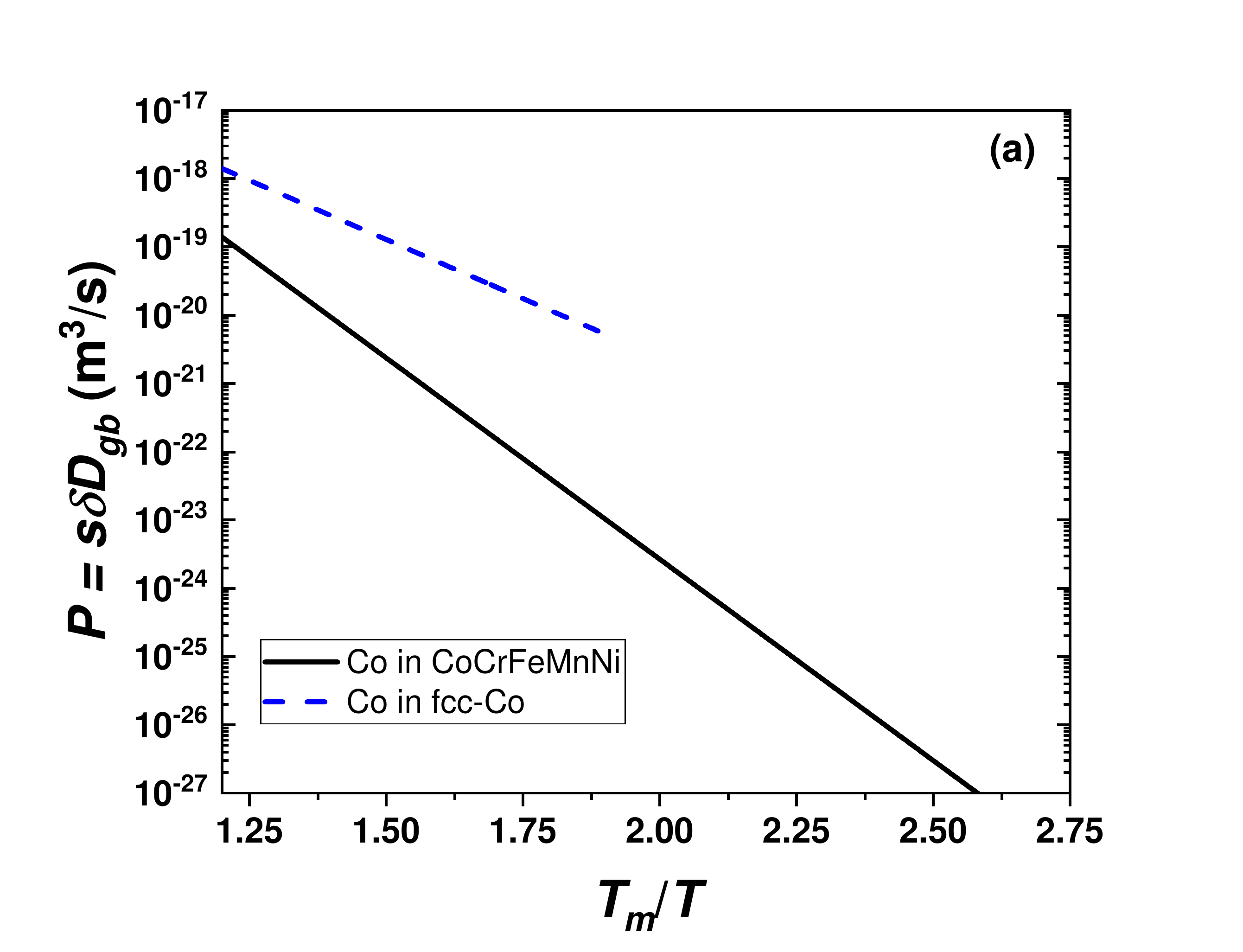}
		\includegraphics[width=0.8\linewidth]{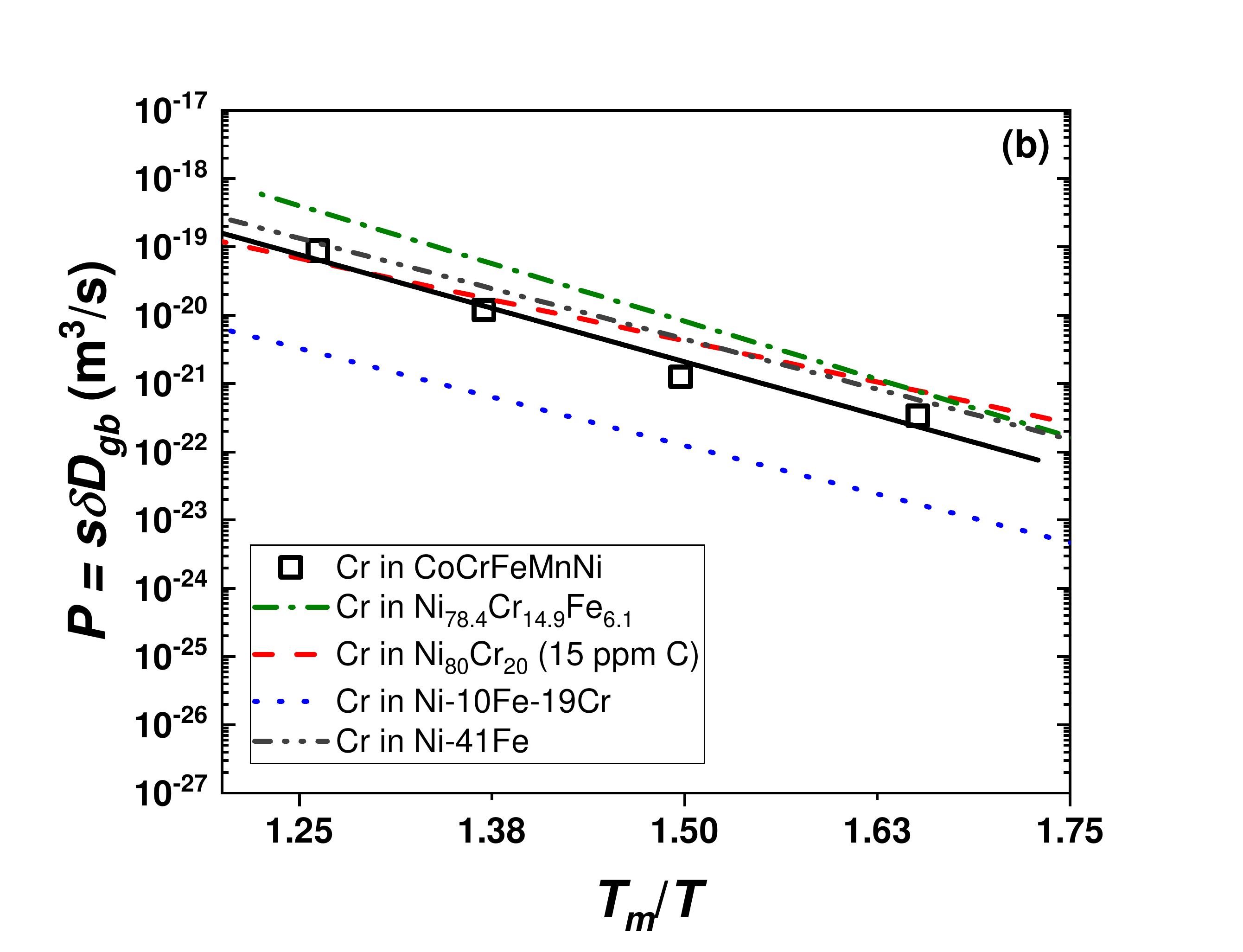}
		\includegraphics[width=0.8\linewidth]{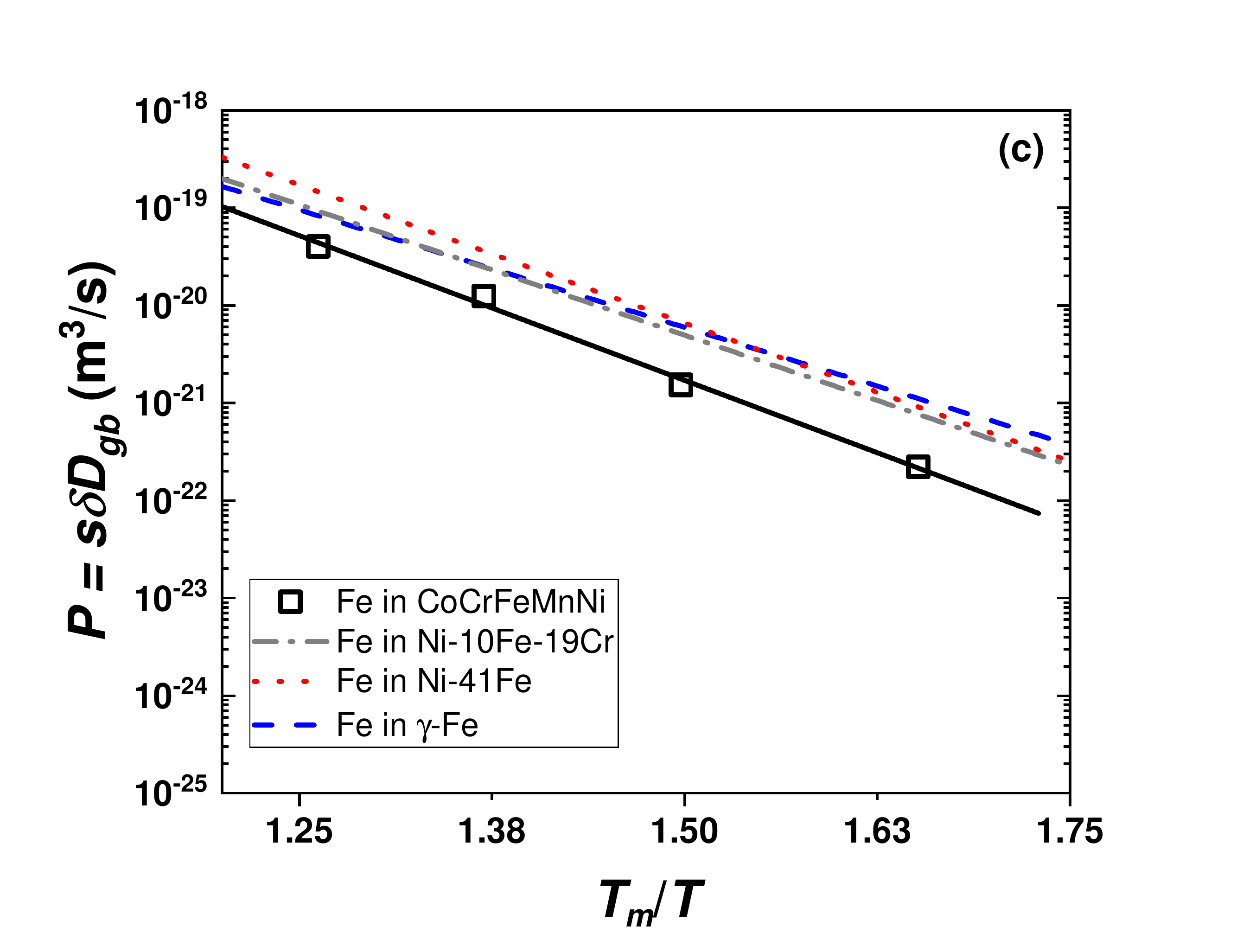}
	\end{center}
	
	\caption{Comparison of Co (a), Cr (b) and Fe (c) GB diffusion rates in the CoCrFeMnNi HEA with those in other fcc matrices on the inverse homologous temperature scale, $T_m/T$. GB diffusion of Co in fcc Co \cite{Co-Co}, of Cr in NiCr \cite{Cr-NiCr, Lacombe},  NiCrFe \cite{Cermak, Cr-NiCrFe-I, Cr-NiCrFe}, and of Fe in $\gamma$-Fe \cite{Fe-Fe, Fe-Fe-G, Fe-Fe-I}, Fe--Ni \cite{Kang1, Kang2}, and FeCrNi \cite{Cermak} is shown.}
	\label{fig:Comp_other_fcc_Tm}
\end{figure}

Figure~\ref{fig:Comp_other_fcc_Tm} compares the GB diffusivities of Co, Cr and Fe in CoCrFeMnNi and other FCC matrices on the inverse homologous temperature scale, $T_m/T$. It should be noted that due to a lack of GB diffusion data for Mn in FCC systems, Mn could not be included. 

The comparison substantiates that the concept of 'sluggish diffusion' cannot be extended to grain boundary diffusion and used as a blanket statement even for FCC matrices. Whereas a gradual retardation of the GB diffusion rates is observed for Fe with an increase of the number of principal alloying elements along a chain $\gamma$-Fe $\rightarrow$ ternary FeCrNi $\rightarrow$ quinary CoCrFeMnNi, this is not the case for Cr diffusion, cf. Figs.~\ref{fig:Comp_other_fcc_Tm}c and b. 
This conclusion correlates generally with that drawn from the analysis of Ni GB diffusivities in Ref.~\cite{GB}. 

A note is due here. Grain boundary diffusion rates depend strongly on the purity of a given base material. While this behavior is well understood for pure metals -- the purer the material, the faster GB self- and impurity diffusion \cite{Surholt, Dasha, Fe-Fe-G, Tok}, the impact of impurities on GB diffusion in multicomponent alloys might be complex and not straightforward \cite{Lacombe}. However, an exact comparison has to include the evaluation of grain size and total amounts of relevant impurities, such as sulfur or carbon \cite{Dasha}.


%

\subsection{Origin of two GB diffusion contributions in the C-type kinetics}

As it is obvious from Fig.~\ref{fig:C-kinetic_profiles}, the penetration profiles at lower temperatures in the C-type kinetic regime exhibit two distinct short-circuit contributions termed as 'slow' and 'fast'. Since volume diffusion is frozen, both these paths correspond to enhanced diffusivities along some defects in the material. The appearance of the two families of short-circuit paths in the CoCrFeMnNi multi-principal element alloy at low temperatures is one of the central findings of the present work. 

Figures~\ref{fig:Arrprofiles}(a)--(d) suggest that the 'slow' paths correspond to random high-angle GBs as they are present in this material at the higher temperatures of the B-type diffusion measurements. Our previous APT investigation \cite{GB} revealed precipitation- and segregation-free high-angle GBs in the CoCrFeMnNi alloy at temperatures above 800~K. The absence of segregation of any constituting element and the same composition in the bulk and at the interfaces suggest that $s \approx 1$ for all elements in the present investigation, at least at $T>800$~K. This conclusion agrees well with the findings in Fig.~\ref{fig:Arrprofiles}, namely the high-temperature $P$-values and low-temperature $\delta \times D_{\gb,1}$-values (if determined assuming $\delta=0.5$~nm) follow the same Arrhenius dependencies over extended temperature interval.

All these facts enable the following important conclusions:
\begin{itemize}
\item the value of 0.5~nm is a very good estimate of the diffusional GB width in the CoCrFeMnNi HEA as for others fcc metals and alloys;
\item the segregation factor for all constituting elements is about unity at $T>800$~K;
\item the general (random) high-angle GBs provide 'slow' contributions to the penetration profiles measured at $T\le700$~K in the C-type kinetics.
\end{itemize}

A single Arrhenius-type temperature dependence for the high-temperature $P$ values and the low-temperature $D_{\gb ,1}$ values excludes dislocations as a potential short-circuit path for 'slow' diffusion. Note that typically the dislocation pipe diffusion coefficients are smaller than $D_{\gb ,1}$ by a factor of 10 to 100 \cite{KMG, Aloke-book}.

The nature of the 'fast' diffusion contribution at lower temperatures in the C-type kinetic measurements has to be clarified. As it was argued, dislocations (randomly distributed or organized in low-angle GBs) cannot provide the 'fast' contribution. 

This reasoning indicates the presence of specific interfaces (others than relaxed high-angle GBs) which could be responsible for the enhanced atomic transport. However, on the micrometer scale, no phase separation or phase decomposition could be observed in the EBSD measurements (Fig.~\ref{fig:EBSD}). Therefore, a correlative microscopy combining APT, Transimisson Kikuchi Diffraction (TKD) and transmission electron microscopy (TEM) investigations has been performed on a CoCrFeMnNi sample annealed at 673~K for 1~week. This annealing treatment mimics the conditions for the C-type diffusion measurements.

\begin{figure*}[ht]
\centering \includegraphics[width=0.9\textwidth]{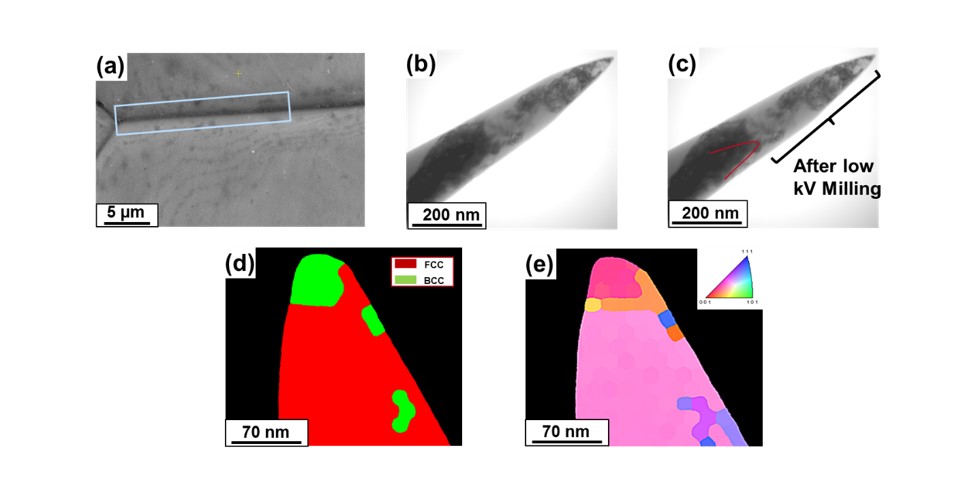}
\caption{(a) SEM image of the grain boundary selected for correlative analysis; (b) BF TEM image indicating the presence of the selected GB along with dislocation sub-structures; (c) BF TEM showing the final tip location enclosing the GB for final (low, 5 kV) milling (d) TKD  Phase map indicating the presence of a bcc precipitate at the apex of the APT tip after low kV milling (e) TKD  inverse pole figure (IPF) map displaying the phase boundaries.}
\label{fig:Corr1}
\end{figure*}

\begin{figure*}[ht]
\centering \includegraphics[width=0.9\textwidth]{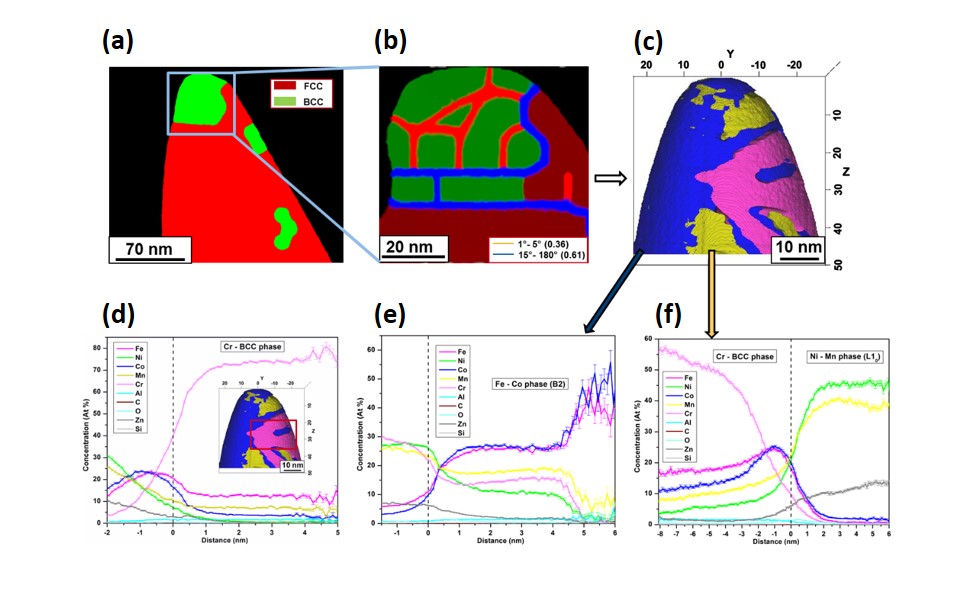}
\caption{(a) TKD phase map of the two phase region enlarged in (b) with image quality (IQ) overlap on the phase map revelaing the GBs (c) the corresponding 3D reconstruction of constituent elements obtained from APT with individual enrichments highlighted using Cr 42 at\%, Co 11 at\%  and Mn 24 at.\% isoconcentration surfaces (d) proximity histogram obtained from the Cr enriched bcc phase (the rectangle shows the Cr rich region in the inset).(e) proximity histogram of the Fe--Co-rich region indicated by an angled arrow from (c) and (f) proximity histogram of the Mn-enriched region indicated by the straight arrow.}
\label{fig:Corr2}
\end{figure*}

Figure~\ref{fig:Corr1}(a) shows a selected region of interest, i.e. a high-angle grain boundary, for site-specific sample preparation denoted by a rectangle. A bright-field (BF) TEM image of the APT specimen is shown in Fig.~\ref{fig:Corr1}(b). It can be noticed that there is a high density of dislocations near the grain boundary and at the bottom of the specimen (dark contrast). In order to measure the grain boundary region in APT, low kV milling was performed to reduce the tip height and position the GB region closer to the apex of the tip as shown in Fig.~\ref{fig:Corr1}(c). For ensuring the presence of the GB and to determine the crystallographic orientation information, TKD was performed on the specimen after low kV milling. Figures~\ref{fig:Corr1}(d) and (e) show the results of the TKD analysis on the APT specimen. The phase map indicates the presence of a nano-scale bcc phase (green colour)  at the apex region and of the matrix fcc phase (red colour) at the bottom of the specimen, while the grain orientation using the inverse pole figure colour coding highlights the presence of a phase boundary at approximately 50 nm from the tip apex. 

In order to correlate the structural information (the results of the TKD analysis) with chemical information, an APT measurement was performed on the same tip that was imaged using TEM and TKD.  Figure~\ref{fig:Corr2}(a) shows the TKD phase map of the APT tip with the apex region (marked rectangle) being enlarged in Fig.~\ref{fig:Corr2}(b) and overlayed with an image quality map (for grain boundary visualization). It can be noticed that the top region of the APT specimen contains a bcc phase with low-angle grain boundaries (misorientations in the range of $1^\circ$ -- $5^\circ$) and several high-angle boundaries (misorientations in the range of $15^\circ$ -- $180^\circ$); their fractions are 0.36 and 0.61, respectively. 

Figure~\ref{fig:Corr2}(c) shows the corresponding elemental reconstruction with the individual enrichments highlighted using the Cr 42 at.\%, Co 11 at.\% and Mn 24 at.\% isoconcentration surfaces. Figure~\ref{fig:Corr2}(d) shows the proximity histogram of the Cr-rich region. It can be noticed from the inset APT reconstruction that the denoted rectangle region corresponds to a bcc phase (from TKD analysis) with Cr concentration of approximately 80~at.\%. This composition is consistent with those observed in larger GB precipitates in the CoCrFeMnNi alloy after prolonged annealing at similar temperatures \cite{Otto, creep}. Figure~\ref{fig:Corr2}(e) shows the proximity histogram of the Fe--Co-rich region. It can be inferred from the proxigrams that the maximum Fe and Co concentrations in the GB nano-precipitates can reach up to 46 at.\% each which indicates formation of a bcc FeCo $\alpha$-phase (probably B2 ordered as it was earlier reported for prolonged annealing treatments \cite{Otto} or under creep conditions \cite{creep} when the particles were large enough for selected area electron diffraction). 

Figure~\ref{fig:Corr2}(e) shows the proximity histogram of the Ni--Mn-rich region. Similar to Fig.~\ref{fig:Corr2} (d) and (e), Mn and Ni concentrations can approach 42 at.\% and 50 at.\% in the enriched region, respectively. Based on this correlative analysis following conclusions can be drawn.

\begin{itemize}
\item  three distinct phases, i.e. Cr-rich bcc, Fe--Co-rich bcc and Ni--Mn-rich $L1_0$ ones, are present along the investigated high-angle GB as nano-scaled precipitates. 
\item Although the composition of the precipitates is similar to what has been observed in the previous literature reports \cite{Otto, creep}, an important difference is that the nano-scale precipitates are seen after annealing at a lower temperature and for a shorter time (673~K for 1~week) with respect to the micro-scale precipitates in Ref.~\cite{Otto}, which were observed after annealing at 773~K for 500~days. 
\end{itemize}

The formation of nano-scale (presumably bcc) Cr-rich and fcc (potentially L1$_0$ ordered) Ni--Mn-rich precipitates at grain boundaries involves appearance the of stress/strain fields and could result in an increased dislocation density in the vicinity of the interface. This is documented in Fig.~\ref{fig:Corr1}(b). In a multi-component environment one may even expect a change of the chemical composition along such dislocation pipes as it was shown in atomistic simulations by Turlo et al. \cite{Turlo-Rup}, termed as linear complexions. 

For the present analysis it is decisive that an increased dislocation density at interfaces or even the formation of GB dislocation networks could  enhance GB diffusion, as it was shown, e.g., for Au diffusion in near $\Sigma5$ GBs \cite{Herzig2}. 

We assume thus that precipitation-induced dislocation density variations modify the GB sub-structure and results in an enhancement of GB diffusion which manifests as 'fast' GB diffusion branch observed in the penetration profiles (Fig.~\ref{fig:C-kinetic_profiles}). The appearance of the two distinct branches suggests that only some fraction of high-angle GBs is covered by the nano-precipitates and provides 'fast' diffusion, while another fraction remains precipitation-free (as it was observed in \cite{GB} using Ni tracer with lower signal-to-background ratio that hindered probably a reliable documentation of the 'fast' diffusion branch). 

Definitely, the concept of two distinct families of interfaces with different diffusivities is an idealization and there is a whole spectrum of interfaces which are probably characterized by continuously varying diffusion coefficients. Still this concept allows a reliable quantification of the diffusion data and agrees qualitatively with the APT results. 

The importance of using $\gamma$-decaying isotopes for measuring GB diffusivities needs to be briefly highlighted here. At low temperatures in the C-type kinetic regime, the corresponding activities in a coarse-grained material are very low and only slightly exceed the background level. The usage of $\gamma$-isotopes and of a stable Ge-detector allows a high sensitivity of detection (the samples were sectioned until distinct $\gamma$-peaks were recorded at the corresponding energies) and a reliable discrimination of the two grain boundary contributions. For example, $^{63}$Ni used in Ref.~\cite{GB} is a $\beta$-decaying isotope and its detection is accompanied by a relatively high background level and, therefore prevented unambiguous detection of the two grain boundary contributions in our previous study. 

The existence of two distinct types of high-angle grain boundaries was advocated according to the present C-type kinetic regime measurements. The diffusion coefficients measured for 'faster' grain boundaries deviate considerably from the Arrhenius dependence established for the B-type kinetic measurements, Fig.~\ref{fig:Arrprofiles}. Thus, such interfaces are not present at higher temperatures of the B-type measurements and some mechanism could have been triggered at lower temperatures (about 600--700~K) affecting the atomic transport along interfaces. It was already observed that grain boundaries can act as starting point for precipitations at low temperatures \cite{Pradeep1, Pradeep2, Raji}. Grain boundary precipitations can induce mechanical stresses and/or be accompanied by the generation of dislocations. 

Fast absolute diffusion rates were already reported in the literature for Ni-based materials at low temperatures. The appearance of such interfaces is e.g. well documented for severely deformed materials \cite{Reg11, Xav, neGBs}. However, in the present case of a coarse-grained and annealed material, the situation is different. 

An impact of precipitation on GB diffusion depends on a number of factors and probably cannot be simply generalized. A retardation of GB diffusion was observed in severely plastically deformed Al--Sc--Zr alloy and it was explained by the formation of coherent Al$_3$(Sc,Zr) precipitates, the appearance of compressive strain around the precipitates and strong segregation of the diffusant (Co atoms in that particular case) to the precipitate/matrix phase boundaries \cite{Vlad}. On the other hand, an enhancement of GB diffusion -- like in the present study -- was observed in a Ni--Cr--Fe alloy at similarly low temperatures of the C-type diffusion measurements due to formation of a secondary dislocation network as a result of stress relaxation induced by carbide precipitation at a fraction of high-angle GBs \cite{Raji}. A similar scenario is proposed to occur in the present equiatomic CoCrFeMnNi alloy, too, and this concept is supported by direct observations via correlated microscopy, Figs.~\ref{fig:Corr1} and \ref{fig:Corr2}.

The appearance of two types of general high-angle GBs with probably similar geometric degrees of freedom and basically different structure and kinetic properties requires a special analysis, especially with respect to a 'selection rule' that forces some boundaries to a phase decomposition while other interfaces -- in fact the majority of them as it follows from the diffusion measurements -- remain untransformed. It is probably not the macroscopic degrees of freedom, i.e. specific misorientation and/or inclination of the interfaces, which trigger the phase decomposition. It was already argued that microscopic degrees of freedom could be responsible for a broad spectrum of different \emph{microstates} of a particular grain boundary \cite{Srolovitz, Frolov}. 

We may speculate that the GB phase decomposition starts at the interface defects like GB disconnections \cite{Srolovitz2, Srolovitz3} and the resulting enhancement of the self-diffusion rate for a particular interface depends on geometric arrangements of the induced secondary dislocation networks. If such network spans over the whole GB plane, the diffusion rate along it increases drastically and is measured in the present experiments as a separate contribution.

If these are stress/strain fields which -- together with the induced secondary dislocation networks -- contribute to the enhanced transport, the diffusion enhancement has to depend on time and relax probably in the course of diffusion annealing treatments.

\subsection{Time dependent measurement in the C-type regime}

\begin{figure}[ht]
\begin{center}
\includegraphics[width=0.49\linewidth]{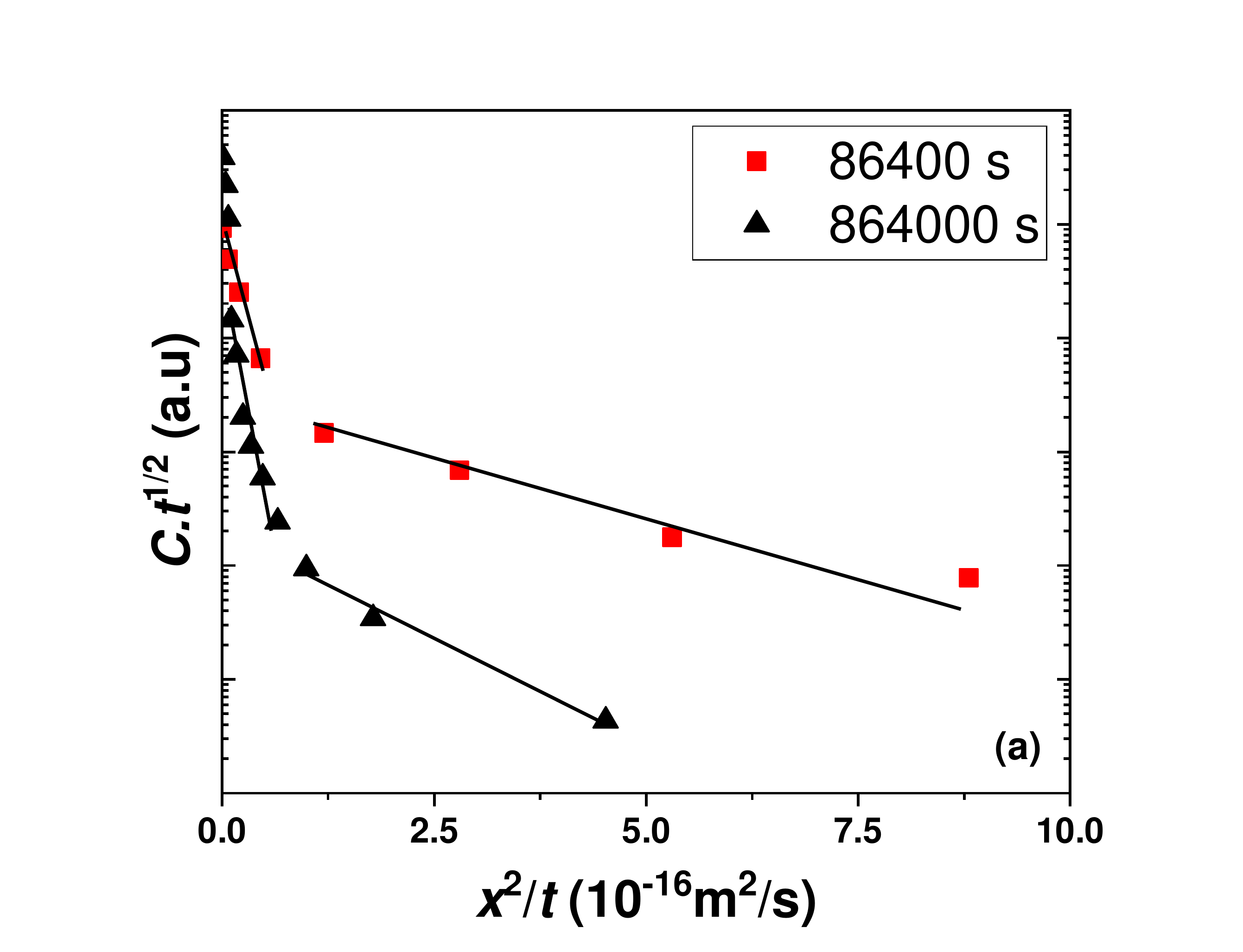}
\includegraphics[width=0.49\linewidth]{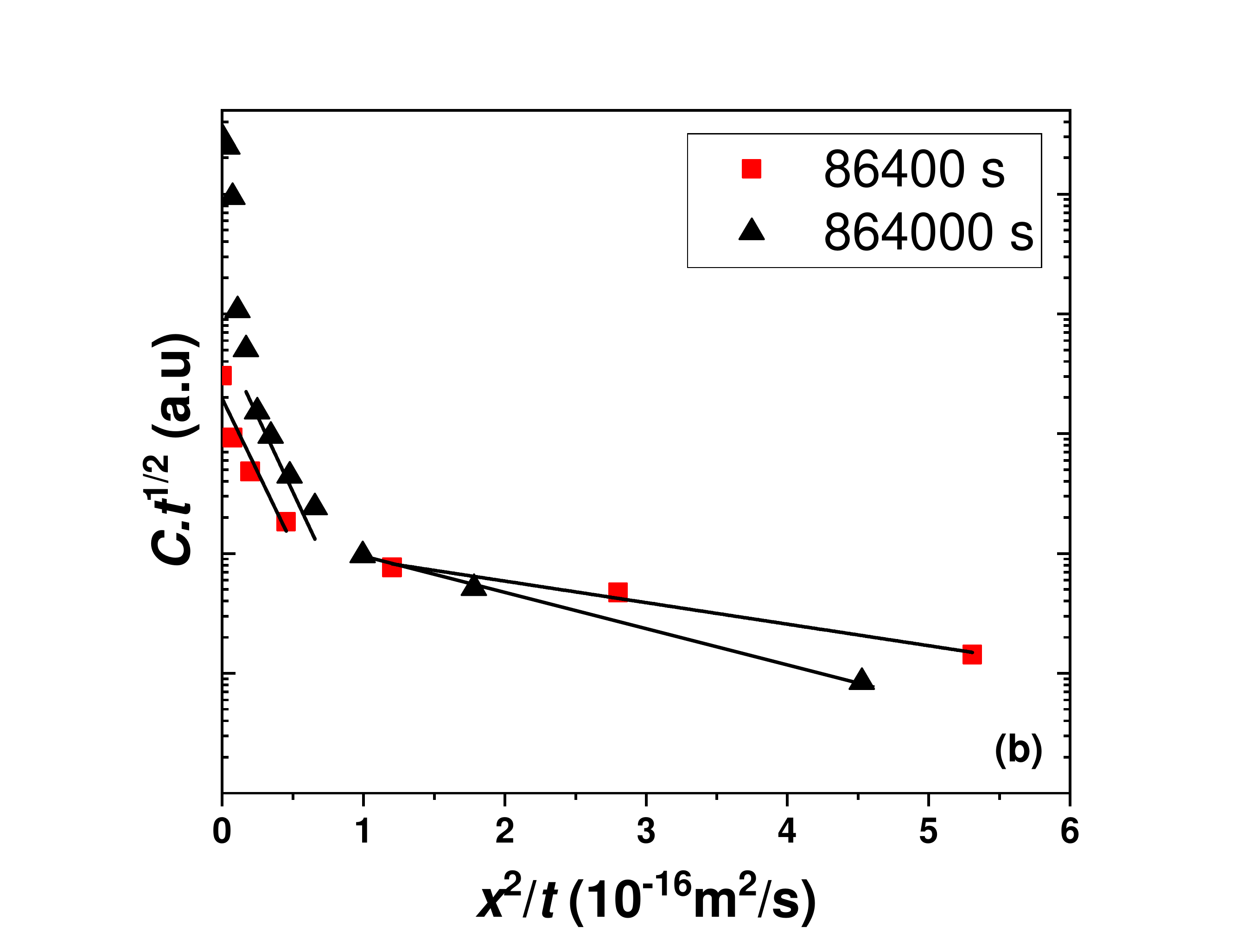}

\includegraphics[width=0.49\linewidth]{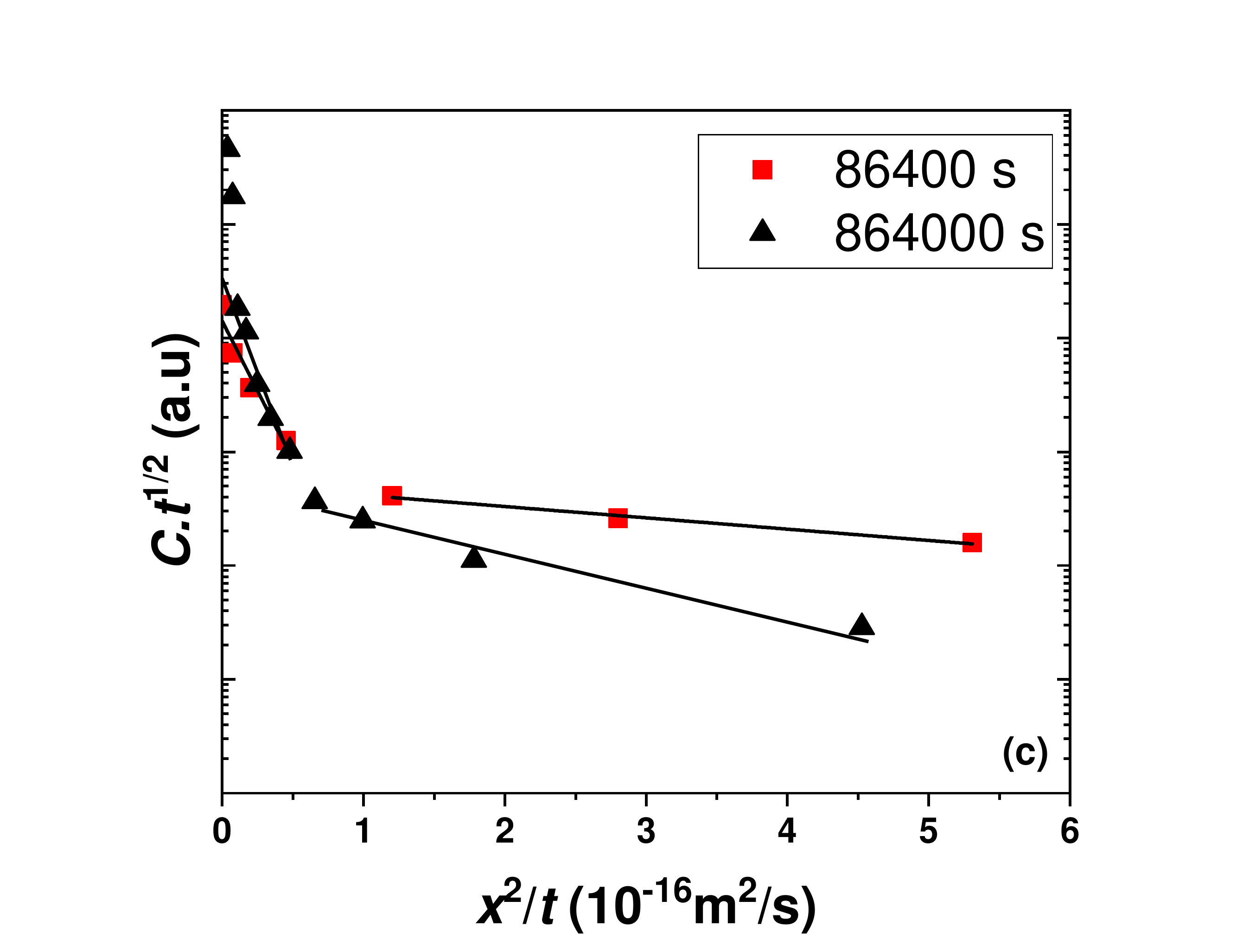}
\includegraphics[width=0.49\linewidth]{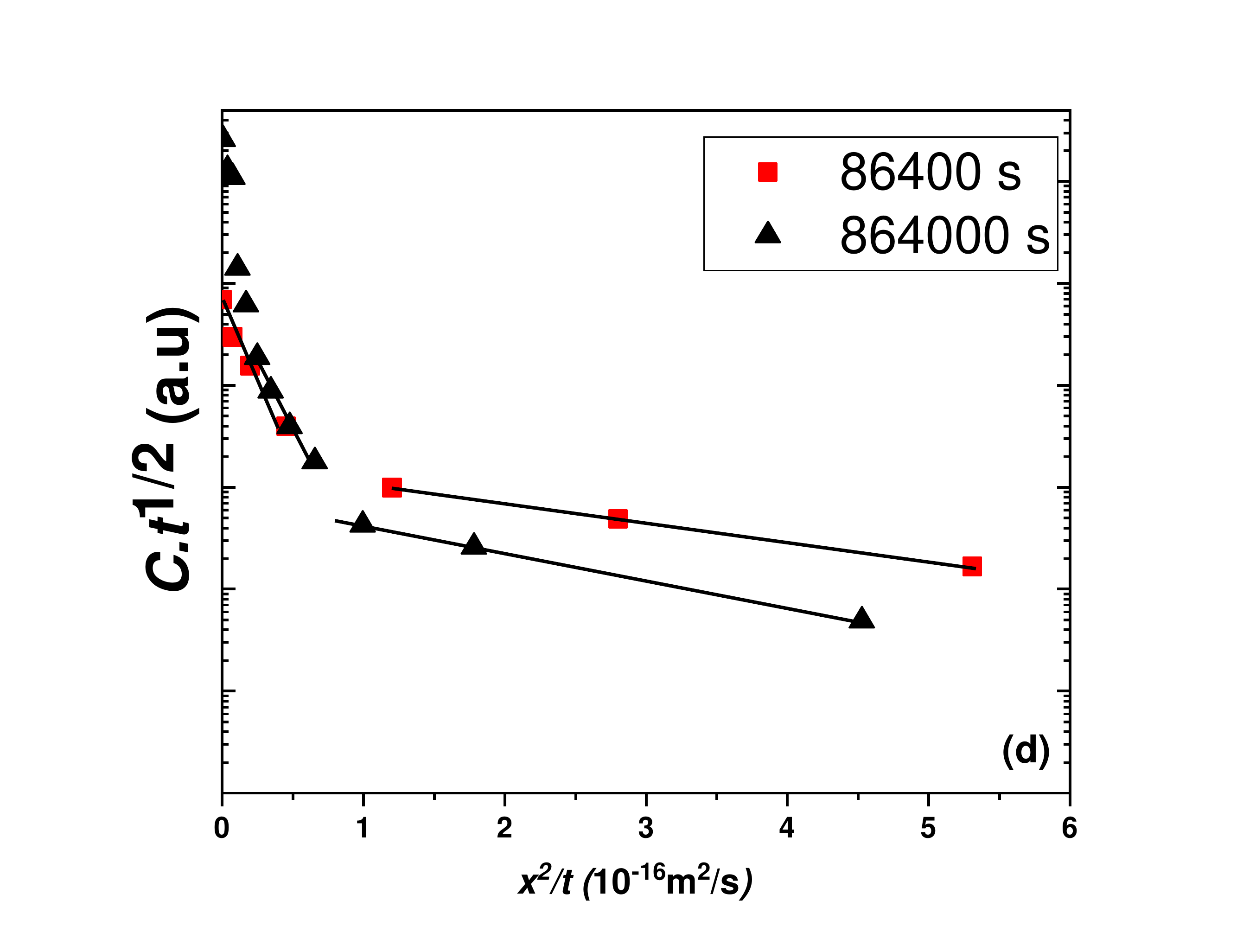}
\end{center}
\caption{Comparison of two experiments at 643~K for $86400$~s and $864000$~s annealing times for Co (a), Cr (b), Mn (c), and Fe (d). The reduced coordinates, $c\cdot t^{1/2}$ vs. $x^2/t$, are used to highlight the deviations from the Gaussian solution with constant diffusion coefficients.}
\label{fig:Comp_c}
\end{figure}

In order to provide further insight into the impact of precipitation on GB diffusion we performed a series of time-dependent measurements at low temperatures. In Fig.~\ref{fig:Comp_c} the penetration profiles measured after annealing at $643$~K for 10 days are compared to those determined after annealing at the same temperature for 1 day (without any pre-annealing at the diffusion temperature after the homogenization annealing at 1373~K for 24~h). The two profiles are compared in reduced coordinates, multiplying the concentrations by $\sqrt{t}$ and dividing depths by $\sqrt{t}$. If the pertinent diffusion coefficient is constant, the Gaussian solution of the diffusion problem has to be valid and the two profiles have to coincide, cf. Eq.~\ref{eq:Gaus}. 

Before comparison of the profiles and the determined diffusion coefficients, see Table~\ref{tab:C-param1}, we highlight that it is almost impossible to apply exactly the same tracer amount to the samples; this explains some deviations of the absolute values of concentrations. Two important conclusions could be drawn:
\begin{itemize}
\item The 'slow' diffusion path is characterized by time-independent diffusion coefficients (different for different elements) and the density of such paths is almost constant;
\item The diffusion coefficients corresponding to the 'fast' paths slightly decrease for longer annealing times.
\end{itemize}
The last point suggests a certain decrease of the diffusion enhancement in the course of the diffusion annealing. Grain boundary diffusion-induced climb of secondary GB dislocations along with a relaxation of the stress/strain fields cause probably the observed deceleration of diffusion for the 'fast' branch.

These results support our interpretation of the nature of 'slow' and 'fast' branches.

\begin{table}[ht] 
\caption{Diffusion coefficients determined after annealing at $T=643$~K for $t = 86400$~s in the C-type kinetics regime.}
\label{tab:C-param1}
\begin{center}
\begin{tabular}{c|cccc}
	\hline
	\multirow{2}{1.1cm}{Tracer} & $\sqrt{D_\vv t}$ & $D_{\gb}$ 'slow' & $D_{\gb}$ 'fast' & \multirow{2}{1cm}{~~~$\alpha$} \\
	& ($10^{-10}$~m) & ($10^{-18}$ m$^2$/s) & ($10^{-17}$ m$^2$/s) & \\
	\hline  & & & & \\[-4pt]
	Co & $0.16$ & $5.89 ^{+3.7}_{-2.5}$ & $7.11 ^{+3.5}_{-3.2}$ &$15 $ \\[4pt]
	Cr & $0.17$ & $5.38^{+ 2.95}_{-1.5}$ & $6.67^{+3.4}_{-3.1}$ &$14 $ \\[4pt]
	Mn & $0.12$ & $6.28^{+2.5}_{-1.36}$ & $7.93 ^{+4.2}_{-3}$ &$20 $ \\[4pt]
	Fe & $0.15$ & $5.4^{+3.6}_{-3.2}$ & $6.62 ^{+5.3}_{-4}$ &$17 $ \\
	\hline
\end{tabular} 
\end{center}
\end{table}

\subsection{Grain boundary energy}

The grain boundary energy, $\gamma_{\gb}$, for pure metals and alloys is an important parameter for a number of nucleation-related phenomena. Borisov et al. \cite{Borisov} suggested a semi-empirical approach which relates the GB energy and self-diffusion rates in pure metals and binary alloys, 
\begin{equation} \label{eq:gb_energyB}
\gamma_{\gb} = \frac{R T}{2 a_0^2 N_a} \ln \bigg( \frac{D_{\gb}}{D_\vv} \bigg).
\end{equation}
where $a_0$ is the lattice constant and $N_a$ the Avogadro number

Later on, Gupta \cite{Gupta} has re-written this expression, introducing explicitly the volume and the grain boundary pre-exponential factors of the diffusion coefficients ($D_{0,\vv}$, $D_{0,\gb}$) and the corresponding activation enthalpies ($\Delta H_\vv$ and $\Delta H_{\gb}$):
\begin{equation} \label{eq:gb_energyG}
\gamma_{\gb} = \frac{RT}{2 a_0^2 N_a} \text{ln} \bigg( \frac{D_{0,\gb}}{D_{0,\vv}} \bigg) + \frac{1}{2 a_0^2 N_A} \big( \Delta H_\vv - \Delta H_{\gb}  \big),
\end{equation}

\begin{figure}[ht]
\centering		\includegraphics[width=0.9\linewidth]{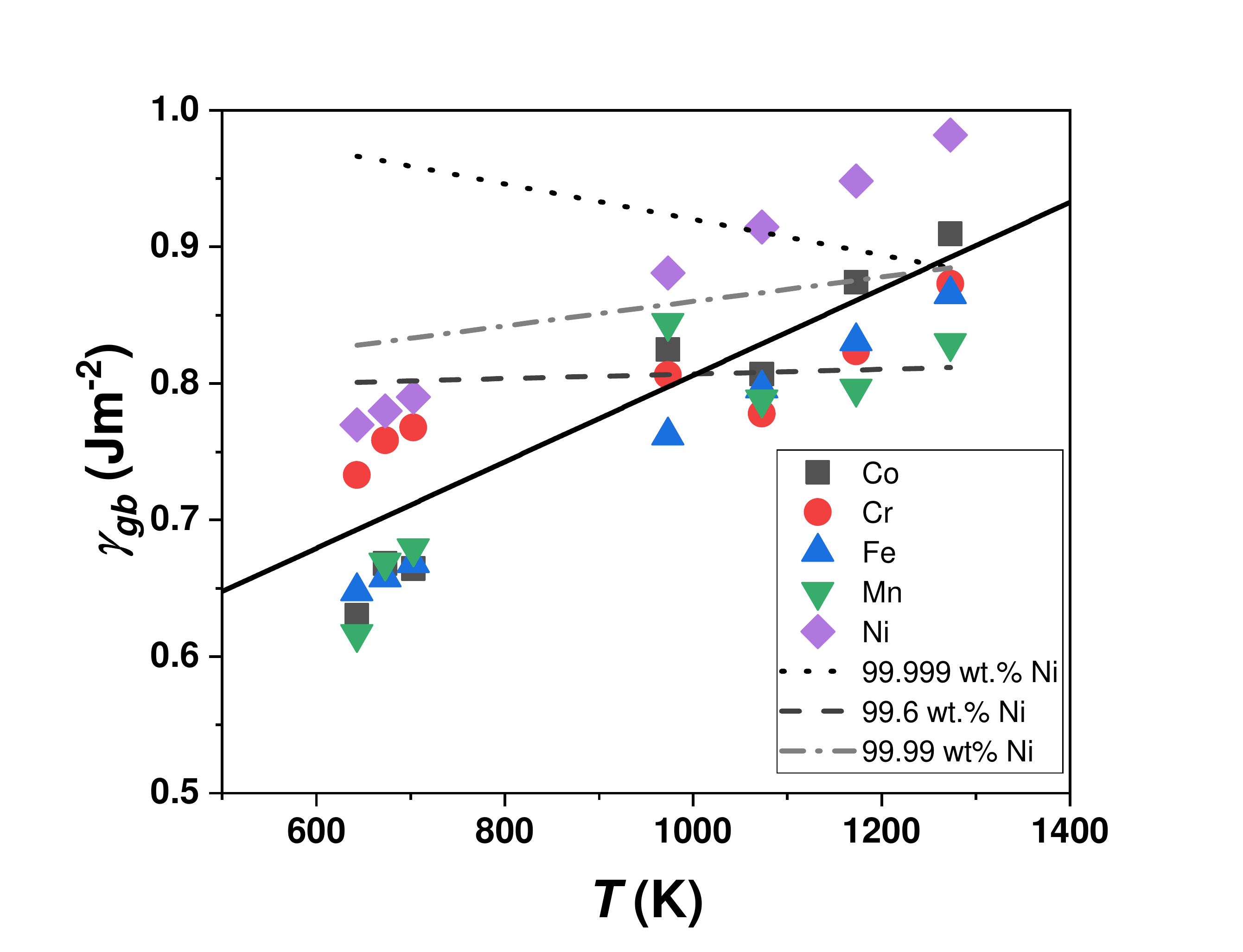}
\caption{Grain boundary energy, $\gamma_\gb$, estimated from the GB diffusion data. The solid black line represents a linear fit to the combined data set for CoCrFeMnNi HEA. For comparison, the GB energies determined for Ni of different purity are drawn by dotted (99.999~wt.\% \cite{Ger}), dash-dotted (99.99~wt.\% \cite{Dasha}) and dashed (99.6~wt.\% \cite{Ger}) lines.}
\label{fig:gb_energy}
\end{figure}

We used these empirical expressions for the present case of multi-component alloys and determined the grain boundary energy using data separately for each constituent. The results are summarized in Fig.~\ref{fig:gb_energy}. The determined energies are relatively similar, which is to be expected as the tracers diffuse along the same grain boundaries. This fact is in favor of an applicability of Borisov's semi-empirical approach to HEAs and the data dispersion might be used as uncertainty range.

Using all data, a common linear temperature dependence of the GB energy can be suggested,
\begin{equation}
\gamma_{\gb} = (0.49 \pm 0.04) + T \cdot (3.17 \pm 0.35)\times 10^{-4} \,\, ({\rm Jm}^{-2}).
\end{equation}
Here the GB energy is given in Jm$^{-2}$ and the temperature in K. The positive slope may be attributed to residual impurities being present in the alloy and segregating towards grain boundaries at low temperatures, decreasing the grain boundary energy. 

In Fig.~\ref{fig:gb_energy}, the grain boundary energy estimated for the CoCrFeMnNi HEA is compared to the grain boundary energies determined previously for $99.6$~wt.\% \cite{Ger}, $99.99$~wt.\% \cite{Dasha} and  $99.9998$~wt.\% \cite{Ger} pure Ni. In high-purity Ni, the GB energy decreases with temperature as it is expected. 
On the other hand, Ni of lower purity reveals positive slopes that was attributed to impurities segregating towards grain boundaries at low temperatures, decreasing their energy \cite{Ger, Dasha}.

\section{Conclusions}

Grain boundary self-diffusion of Co, Cr, Fe and Mn was measured in chemically homogeneous and single phase (solid solution) CoCrFeMnNi HEA using the radiotracer method. The measurements were performed in both, the B- and C-type kinetic regimes of GB diffusion. 

At elevated temperatures of the B-type kinetic measurements, above 700~K, tracer diffusion along general (random) high-angle GBs was determined. An APT investigation in our previous study on Ni GB diffusion in the same alloy revealed a chemical homogeneity at the interfaces, i.e. an absence of element segregation and/or precipitation \cite{GB}.

The penetration profiles measured in the C-type kinetic regime (643--703~K) revealed two short-circuit contributions to the GB diffusion transport, termed as 'slow' and 'fast' branches. The atom probe tomography substantiated the formation of Ni-Mn-rich and Cr-rich nano-precipitates at some inspected high-angle GBs. Transmission electron microscopy results documented an enhanced dislocation density in the vicinity of such GBs. These dislocations were suggested to provide the observed enhancement of GB diffusion at the corresponding interfaces, which were associated with the 'fast' branch. The 'slow' branch corresponds to diffusion along the high-angle GBs which have not underwent precipitation or decomposition.

The grain boundary triple product, $P$, determined in the B-type kinetic regime and the product $\delta \times D_{\gb}$ (using the values $D_{\gb,1}$ determined for the 'slow' branch) follow consistently single Arrhenius dependencies for all elements. This fact substantiates that the product of the segregation factor and the GB width, $s\times \delta$, is about 0.5~nm. We interpret this result as the absence of element segregation at higher temperatures of the B-type diffusion measurements, $s\approx1$. Simultaneously the GB width in HEAs can be estimated as 0.5~nm.

Very similar activation enthalpies of GB diffusion in CoCrFeMnNi are determined for Co, Cr, Fe and Mn, of about 180~kJ/mol. 

The grain boundary diffusivities obtained in the present work were used to estimate the GB energy in CoCrFeMnNi HEA, which was found to approach 0.49~Jm$^{-2}$ at 0~K and it increases with increasing temperature.

\vspace{0.5cm}
\noindent{\bf Acknowledgments}. Financial support of the German Science Foundation (DFG) via research projects DI 1419/13-2 and WI 1899/32-1 is acknowledged. KGP is grateful for the financial support of Max-Planck Gesellschaft through Max-Planck-India partner group. KGP and KG acknowledge the use of National Facility for Atom Probe Tomography (NFAPT) for correlative microscopy studies.

\end{document}